\documentclass[12pt]{oxarticle}
\pdfoutput=1

\usepackage{graphicx, float, array, xspace, amscd, amsmath, amsthm, amssymb, latexsym}
\usepackage{amsfonts}
\usepackage[bbgreekl]{mathbbol}

\DeclareSymbolFontAlphabet{\mathbb}{AMSb}
\DeclareSymbolFontAlphabet{\mathbbl}{bbold}

\usepackage[sort&compress, comma, square, numbers]{natbib}
\usepackage[all,cmtip]{xy}

\usepackage{color}
\definecolor{MyDarkBlue}{rgb}{0.15,0.25,0.45}

\let\SS=\S 

\def\starb{\,{\overline{ \star}\,}}

\newcommand{\eb}{{\overline{\eta}}}
\renewcommand{\sb}{{\overline{\sigma}}}

\newcommand{\kb}{{\overline{ k}}}
\newcommand{\rb}{{\overline{ r}}}

\newcommand{\Ob}{{\overline{ \Omega}}}

\newcommand{\w}{{\,\wedge\,}}

\newcommand{\pre}[1]{{}^{(#1)}}

\newcommand{\Le}{{\mathfrak{e}}}

\newcommand{\Lg}{{\mathfrak{g}}}

\newcommand{\half}{\frac{1}{2}}
\newcommand{\qrt}{\frac{1}{4}}

\def\rep#1{{{\boldsymbol{#1}}}}
\def\brep#1{{{\overline{\boldsymbol{#1}}}}}

\def\CS{{\text{CS}}}

\newcommand{\bb}{{\overline\beta}}

\renewcommand{\a}{\alpha}
\renewcommand{\b}{\beta}
\newcommand{\g}{\gamma}
\renewcommand{\d}{\delta}\newcommand{\D}{\Delta}
\newcommand{\e}{\epsilon}
\newcommand{\z}{\zeta}

\newcommand{\Th}{\Theta}\newcommand{\vth}{\vartheta}

\renewcommand{\k}{\kappa}
\renewcommand{\l}{\lambda}\renewcommand{\L}{\Lambda}
\newcommand{\m}{\mu}
\newcommand{\n}{\nu}
\newcommand{\x}{\xi}\newcommand{\X}{\Xi}

\renewcommand{\P}{\Pi}
\renewcommand{\r}{\rho}
\newcommand{\s}{\sigma}\renewcommand{\S}{\Sigma}
\renewcommand{\t}{\tau}
\newcommand{\U}{\Upsilon}
\newcommand{\ph}{\phi}\newcommand{\Ph}{\Phi}\newcommand{\vph}{\varphi}
\newcommand{\ch}{\chi}
\newcommand{\ps}{\psi}\newcommand{\Ps}{\Psi}
\renewcommand{\o}{\omega}\renewcommand{\O}{\Omega}

\DeclareFontFamily{OT1}{pzc}{}
\DeclareFontShape{OT1}{pzc}{m}{it}{<-> s * [1.200] pzcmi7t}{}
\DeclareMathAlphabet{\mathpzc}{OT1}{pzc}{m}{it}

\newcommand{\cA}{\mathcal{A}}\newcommand{\ccA}{\mathpzc A}
\newcommand{\ccB}{\mathpzc B}

\newcommand{\ccD}{\mathpzc D}
\newcommand{\ccE}{\mathpzc E}
\newcommand{\cF}{\mathcal{F}}
\newcommand{\cG}{\mathcal{G}}
\newcommand{\cH}{\mathcal{H}}\newcommand{\ccH}{\mathpzc H}

\newcommand{\cL}{\mathcal{L}}
\newcommand{\cM}{\mathcal{M}}\newcommand{\ccm}{\mathpzc m}

\newcommand{\cO}{\mathcal{O}}

\newcommand{\cQ}{\mathcal{Q}}
\newcommand{\cR}{\mathcal{R}}

\newcommand{\ccT}{\mathpzc T}

\newcommand{\ccX}{\mathpzc X}

\DeclareFontFamily{U}{bbold}{}
\DeclareFontShape{U}{bbold}{m}{n}
 {  <-5.5> s*[1.05] bbold5
    <5.5-6.5> s*[1.05] bbold6
    <6.5-7.5> s*[1.05] bbold7
    <7.5-8.5> s*[1.05] bbold8
    <8.5-9.5> s*[1.05] bbold9
    <9.5-11.5> s*[1.05] bbold10
    <11.5-16> s*[1.05] bbold12
    <16-> s*[1.05] bbold17
 }{}

\newcommand{\IA}{\mathbbl{A}}

\newcommand{\IC}{\mathbbl{C}}
\newcommand{\Id}{\mathbbl{d}}

\newcommand{\IF}{\mathbbl{F}}
\newcommand{\IG}{\mathbbl{G}}

\newcommand{\IR}{\mathbbl{R}}

\newcommand{\IZ}{\mathbbl{Z}}

\newcommand{\ITheta}{\mathbbl{\Theta}}

\font\eightrm=cmr8 at 8pt
\font\csc=cmcsc10

\newcommand{\beq}{\begin{equation}}
\newcommand{\eeq}{\end{equation}}
\newcommand{\beqnn}{\begin{equation*}}
\newcommand{\eeqnn}{\end{equation*}}
\newcommand{\bea}{\begin{eqnarray}}
\newcommand{\eea}{\end{eqnarray}}
\newcommand{\bean}{\begin{eqnarray*}}
\newcommand{\eean}{\end{eqnarray*}}

\newcommand{\sref}[1]{\SS\ref{#1}}

\newcommand{\pd}[2]{\frac{\partial #1}{\partial #2}}
\newcommand{\diff}[1]{\frac{\partial}{\partial #1}}

\newcommand{\norm}[1]{\left\| #1\right\|}
\newcommand{\ee}{\text{e}}
\newcommand{\ii}{\text{i}}

\newcommand{\place}[3]{\vbox to0pt{\kern-\parskip\kern-7pt
                             \kern-#2truein\hbox{\kern#1truein #3}
                             \vss}\nointerlineskip}
\newcommand{\smallfrac}[2]{\frac{\scriptstyle #1}{\scriptstyle #2}}

\DeclareFontFamily{U}{wncy}{}
\DeclareFontShape{U}{wncy}{m}{n}{<->wncyr10}{}
\DeclareSymbolFont{mcy}{U}{wncy}{m}{n}
\DeclareMathSymbol{\sha}{\mathord}{mcy}{"58}

\newcommand{\eu}[1]{{\mathfrak #1}}

\newcommand{\capt}[3]{\parbox{#1}{\renewcommand{\baselinestretch}{1.0}
                                                           \caption{\label{#2}\small\it #3}}}

\newcommand{\del}{\partial}
\newcommand{\delb}{\overline{\partial}}

\newcommand{\jb}{{\bar\jmath}}

\newcommand{\nb}{{\bar\n}}
\newcommand{\mb}{{\bar\m}}

\newcommand{\delbA}{\delb_{{}\hskip-2.8pt\cA}}

\newcommand{\A}{\cA}

\newcommand{\End}{\text{End}}
\newcommand{\dd}{\text{d}}
\newcommand{\Hend}{H^1(\ccX,\text{End}\,\ccE)}

\newcommand{\cym}{Calabi-Yau manifold\xspace}

\newcommand{\K}{K\"ahler\xspace}

\newcommand{\cs}{\text{\tiny CS}}

\newcommand{\YM}{\text{\tiny YM}}

\def\ker{{\rm ker ~}}

\newcommand{\tr}{\,\text{Tr}\,}

\newcommand{\tb}{{\overline{\tau}}}

\newcommand{\ap}{{\a^{\backprime}\,}}
\renewcommand{\sb}{{\overline{\sigma}}}
\renewcommand{\rb}{{\overline{\rho}}}

\makeatletter
\g@addto@macro\bfseries{\boldmath}
\makeatother

\renewcommand{\baselinestretch}{1.1}
\numberwithin{equation}{section}
\proofmodefalse
\begin{document}
\pagestyle{empty}

\ifproofmode\underline{\underline{\Large Working notes. Not for circulation.}}\else{}\fi

\begin{center}
\null\vskip0.2in
{\Huge A Metric for Heterotic Moduli\\[0.5in]}
{\csc Philip Candelas$^{*\,1}$, Xenia de la Ossa$^{*\,2}$\\
and\\
Jock McOrist$^{\dagger \,3}$\\[0.5in]}
{\it $^*$Mathematical Institute\hphantom{$^1$}\\
University of Oxford\\
Andrew Wiles Building\\
Woodstock Road, Radcliffe Observatory Quarter\\
Oxford, OX2 6GG, UK\\[3ex]
$^\dagger$Department of Mathematics\hphantom{$^2$}\\
University of Surrey\\
Guildford, GU2 7XH, UK\\
}
\vfill
{\bf Abstract\\}
\end{center}
\vspace*{-5pt}
Heterotic vacua of string theory are realised, at large radius, by a compact threefold with vanishing first Chern class together with a choice of stable holomorphic vector bundle. These form a wide class of potentially realistic four-dimensional vacua of string theory. Despite all their phenomenological promise, there is little understanding of the metric on the moduli space of these. What is sought is the analogue of special geometry for these vacua. The metric on the moduli space is important in phenomenology as it normalises D-terms and Yukawa couplings. It is also of interest in mathematics, since it generalises the metric, first found by Kobayashi, on the space of gauge field connections, to a more general context. Here we construct this metric, correct to first order in $\ap$, in two ways: first by postulating a metric that is invariant under background gauge transformations of the gauge field, and also by dimensionally reducing heterotic supergravity. These methods agree and the resulting metric is \K, as is required by supersymmetry. Checking  the metric is  \K is  intricate and the anomaly cancellation equation for the $H$ field plays an essential role. The \K potential nevertheless takes a remarkably simple form: it is the  \K potential of special geometry with the \K form replaced by the $\ap$-corrected hermitian~form. 

\newpage
{\renewcommand{\baselinestretch}{1.4}\tableofcontents}
\newpage
\setcounter{page}{1}
\pagestyle{plain}
\section{Introduction}
\vskip-10pt
Our aim is to describe the metric on the space of heterotic vacua. This parameter space geometry, which we term {\it heterotic geometry}, is the generalisation of the special geometry of type II string theory. The heterotic vacua of concern here derive from compactifying heterotic string theory, at large radius, on 
$\IR^{3,1}{\times} \ccX$, where $\ccX$ is a smooth complex threefold with vanishing first Chern-class that is endowed with a holomorphic vector bundle $\ccE$, that has a connection satisfying the Hermitian--Yang--Mills (HYM) equation and a gauge invariant three-form $H$. These quantities satisfy an anomaly condition, that will be discussed shortly. These vacua are of physical interest since, at low-energies, they realise quasi-realistic four-dimensional theories of relevance to observable particle physics. 

The geometry of the moduli spaces of these vacua are also of mathematical interest. The metric on the local moduli space of holomorphic hermitian Yang-Mills (HYM) bundles $\ccE$ constructed over an arbitrary but fixed complex manifold goes back to Kobayashi~\cite{Kobayashi:1987} and collaborators. By constructing local coordinates in the spirit of Kodaira-Spencer, this metric was shown to be \K by Itoh \cite{Itoh:1988}. However, the restriction to a fixed CY manifold is artificial from the point of view of string theory:  the moduli space includes deformations of $\ccX$, the gauge-invariant three-form $H$ and the vector bundle $\ccE$ simultaneously.  We call the triple $(\ccX, \ccE, H)$ a {\it heterotic structure}. The present work is complementary to a series of papers \cite{delaOssa:2014cia, delaOssa:2014msa,delaOssa:2015maa,Anderson:2014xha}, which describe this heterotic structure and identify the moduli of the vacua with certain cohomology groups. 
 
A key lesson that arises in our analysis, as in the papers that have just been cited, is that even when the compactification is defined with $\ccX$ a manifold that is CY in the limit $\ap\to 0$, the torsion defined by the 
$H$-field, which arises already at $\cO(\ap)$, is a crucial part of the story. Indeed, this correction is intricately related to the Hermitian form $\o$ for $\ccX$ through the supersymmetry relation
\beq
 H ~=~ \dd^c \o~,\quad \dd^c \o = J^m\del_m \o -( \dd J^m)\o_{mn} \dd x^n~, 
\label{eq:SusyRelation0}\eeq
together with the anomaly relation
\beq
\dd H =- \frac{\ap}{4} \left( \tr F^2 - \tr R^2\, \right)~,
\label{eq:Anomaly0}\eeq
where, in the first of these equations, $J^m{\,=\,}J_k{}^m \dd x^k$ is a one-form constructed from the complex structure, $\o$ is the hermitian form of $\ccX$ and in the second $F$ is the gauge field strength while $R$ denotes the curvature two-form. There are two immediate observations: firstly, if we are genuinely in the heterotic setting with $F\neq R$, then, generically, $H$ cannot vanish, so 
$\o{\,=\,}\ii g_{\m\nb}\dd x^\m \dd x^\nb$ cannot be a \K-form (since such a form would satisfy 
$\del\o{\,=\,}\delb\o{\,=\,}0$). Secondly, taking these two relations together, we see that variations of the hermitian form $\o$ are related to variations of the B-field {\em and\/} the background gauge field and so the vector bundle~$\ccE$. From a mathematical perspective these are unexpected and new ingredients, but prove crucial in constructing a moduli space that is \K, as required by supersymmetry. 

A consequence of these observations is that the naive expectation, that the parameters of heterotic compactifications could be thought of as corresponding to the cohomology groups
$H^1(\ccT_\ccX)\oplus H^1(\ccT^*_\ccX)\oplus H^1(\text{End}\ccT_\ccX)$, where $\ccT_\ccX$ denotes the tangent bundle of $\ccX$, is far from the case. Even without the considering the consequences of the anomaly and supersymmetry relations, the Atiyah constraint \cite{Atiyah:1955}, which has long been known, was a warning that the complex structure parameters of the manifold cannot be varied arbitrarily, since they are constrained by the requirement that the bundle remain holomorphic. 
We will review this in the following, but the essential point is that a bundle is holomorphic if and only if the Yang-Mills field strength $F$ satisfies
\beqnn
F^{(2,0)}~=~F^{(0,2)}~=~0\,,
\eeqnn 
and so $F$ must be precisely of type $(1,1)$. However, under a variation of complex structure a form of type $(1,1)$ will, in general, acquire parts of type $(2,0)$ and $(0,2)$. The condition that this does {\em not\/} happen is that 
\beqnn
\D^\m\wedge F_\m~=~\delb_\A (\d \A)~, 
\eeqnn
for some $\d \A$. In other words $\D^\m\w F_\m {\,=\,} 0$, in cohomology.  Our notation is that  $\D$ is a (0,1)-form with values in $\ccT^{1,0}_\ccX$, that corresponds to the variation of the complex structure, and 
$F_\m{\,=\,}F_{\m\bar\n}\dd x^{\bar\n}$ is a (0,1)-form constructed from $F$, $\cA$ is the $(0,1)$-part of the gauge potential and $\delb_\cA$ is the $(0,1)$-part of a gauge covariant derivative. In this equation we have written the exterior multiplication explicitly. In the following we generally understand the multiplication of forms to be exterior multiplication, writing `$\wedge$' or $\otimes$ only where confusion may arise.  We view the equation above as a restriction on the allowed values of the complex structure parameters of~$\ccX$. 
 
On its own, the Atiyah constraint is perhaps not so serious, since it can be viewed as merely reducing the number of allowed complex structure parameters. This reduction of complex structure parameters was studied in the context of heterotic supergravity in several examples in \cite{Anderson:2010mh,Anderson:2011ty}. The mixing of the complex structure, \K and bundle deformations required by the anomaly and supersymmetry constraints is, however, far more serious. One might even doubt that there are any consistent variations. Fortunately, the conclusion of the analysis of  
\cite{delaOssa:2014cia} and \cite{Anderson:2014xha} is that the full system of constraints, which now involves the complex structure, bundle and \K parameters, has a character analogous to requiring that the combined variation of the manifold and bundle preserve the condition that the bundle remain holomorphic. So it seems probable that generically there will, in fact, be solutions. This is consistent with the analysis of  $(0,2)$ sigma models and their corresponding conformal field theories, see for example \cite{Melnikov:2012hk,McOrist:2010ae} for some recent reviews. 

To start, it is instructive to compare heterotic geometry with special geometry. Special geometry, familiar from type II string theory with $N{\,=\,}2$ supersymmetry, exists in heterotic string theory with $N {\,=\,} 1$ spacetime supersymmetry when $\ccE$ is identified with the tangent bundle $\ccT_\ccX$. A connection for 
$\ccE$, denoted $A$ is identified with the spin connection of $\ccT_\ccX$, and so the Bianchi identity is trivially satisfied with $\ccX$  a \cym. The moduli space splits as a direct product of complex structure deformations and complexified \K deformations. The metrics on these moduli spaces derive from the Weil-Peterson inner product of harmonic representatives of cohomology groups $H^1(\ccX,\ccT_\ccX)$ and $H^1(\ccX,\ccT^{*}_\ccX)$.    It is important that the holomorphic 3-form $\O$ is a section of a line bundle over the complex structure moduli space, and the underlying space is \K, as this allows the metrics to be rewritten in a way that depends only on the cohomology classes of the variations. As a consequence, it is straightforward to write down the corresponding \K potential
\beq
K = -\log \left(\ii \int \O \wedge \Ob\right) - \log\left( \frac{4}{3} \int \o^3\right)~.
\label{eq:SGKahlerPotential}\eeq

In heterotic geometry, we have a family of heterotic structures $(\ccX, \ccE, H)$ over a moduli space $\cM$. Although the underlying manifold $\ccX$ need not be \K, the moduli space is  required to be \K by supersymmetry. The challenge is to identify the corresponding \K metric and potential. Motivated by special geometry, one might guess that the metric is the natural inner product on the deformations making up the heterotic structure. Remarkably, this is not  far from the final answer. The challenge is to identify exactly the appropriate fields and  deformations. Doing so is subtle since, in general, there is no invariant distinction between complex structure, the \K deformations of $\ccX$, and deformations of the connection for~$\ccE$. Rather, one needs to study deformations of the total space of the vector bundle, while also disentangling the relationships between these fields as dictated by supersymmetry \eqref{eq:SusyRelation0} and the anomaly equation \eqref{eq:Anomaly0}. The appropriate symmetries need to be satisfied; in special geometry the metric for the complex structure deformations needs to respect the fact that there is an extra gauge invariance owing to the fact that the holomorphic three-form is a section of a line bundle. Analogously, the metric here needs to respect a background gauge invariance. Finally the metric needs to be \K, and showing this is so turns out to be a non-trivial consequence of \eqref{eq:SusyRelation0}.  It should be emphasised that although one can consider an important case where $\ccE$ is a deformation of $\ccT$, our results are not tied this, and apply for a general heterotic vacuum.  

In limiting cases, we can get a sense for how this might work. When $\ccX$ is fixed , the parameter space reduces to deformations of the connection for $\ccE$ preserving the HYM condition and the holomorphicity of the bundle. For a bundle with a structure group that is a subgroup of a unitary group the gauge potential is antihermitean and we may write
\beq
A~=~\cA - \cA^\dag~,
\notag\eeq
where $\cA$ is the $(0,1)$ part. Deformations, $\d\cA$, of the connection $\A$ preserving the holomorphicity and HYM conditions of $\ccE$, are elements of the cohomology group $\Hend$. The  metric on this parameter space, first studied by Kobayashi and Itoh, derives from the natural inner product on this cohomology group,
\beq
\dd s^2~=~\frac{\ii\ap}{8 V} \int \o^2 \tr\big(\d\cA\,\d\cA^\dag\big)~,
\label{KobayashiMetric}\eeq
where $V$ is the volume of $\ccX$ and $\o$ denotes the \K-form. The 
prefactor, $\ap/8V$, has been chosen with a certain prescience. It is, of course, a constant for each $\ccX$, but depends on the \K parameters and so is important for us here.

Using the Kuranishi map, Itoh, {\it loc.\ cit.}, constructed a set of coordinates $w^i$ on a slice of  deformations of $\d\cA$ defined by imposing a background gauge fixing. In the language we will use here, the local coordinates amount to a parameterisation $\d\cA= \d w^i \a_i$, where $\a_i$ is a harmonic basis for $\Hend$. Itoh used this coordinate system to write \eqref{KobayashiMetric} as $\dd s^2 = G_{i\jb}\, \dd w^i \dd w^{\jb}$ where 
\beqnn
G_{i\jb}~=~\frac{\ii\ap}{8 V} \int \o^2 \tr\big(\a_i\, \a_\jb^\dag\big)~,
\eeqnn
and then showed that this metric is \K by checking that  $\del_{[i} G_{j]\kb} = 0$.  Itoh assumes that 
$\d A$ satisfies a background field gauge condition and is appropriately harmonic. However, it must be the case that  the metric \eqref{KobayashiMetric} and its \K properties are independent of the choice of gauge for $\A$. A key part of our discussion  will be this demand. 

What we do here, is starting with that parameter space, introduce a set of local coordinates for it $y^M$, and a holomorphic structure so that $y^M =( y^\x, y^\eb)$. With respect to these coordinates we construct gauge covariant derivatives $\ccD_\x$ and $\ccD_\eb$, which when acting on a form respect both its gauge properties and its holomorphic type. This is the natural generalisation of the derivative of $\O$, which respects the fact is that it is a section of a line bundle over the complex structure moduli space. This certainly includes the gauge field $\ccD_\x \A$, but also other forms such as $\ccD_\x \o$. We term these {\it holotypical derivatives}. At order $\ap$, this is complicated by the $H$-field, which as it is defined in terms of Chern-Simons forms for $A$ and the spin connection $\Th$, requires a gauge-dependent B-field. So we construct a holotypical derivative $\ccD_\x B$ which respects this gauge transformation law. The holotypical derivatives then serve as a basis for deformations of the fields defining the heterotic structure $(\ccX,\ccE,H)$. The deformations are related by the supersymmetry condition \eqref{eq:SusyRelation0} together with the anomaly condition \eqref{eq:Anomaly0} and one needs to carefully take this into account. For example, we find that the holotypical derivative of the B-field $\ccD_\x B$ is not an independent deformation and is slaved to deformations of $\ccD_\x \A$ and $\ccD_\x \o$.

Given a basis of deformations we can construct a \K metric, which is the natural inner product of these forms. The \K property follows after taking into account the relations implied by supersymmetry. When written in the basis defined via the holotypical derivatives, the final answer is straightforward to state. The metric is defined  by a \K potential
\beq
 K ~=~- \log\left( \ii \!\int\! \O\, \Ob \right) - \log\left( \frac{4}{3} \int \o^3\right),
\label{eq:KahlerPot0}\eeq
where $\O$ is the holomorphic $(3,0)$-form and now $\o$ the $\ap$-corrected hermitian form.  Remarkably it is identical in form to the special geometry \K potential \eqref{eq:SGKahlerPotential}, and remains quasi-topological, despite the fact that the bundle $\ccE$ does not enter this expression explicitly. The dependence of the bundle parameters arises through the mixing of fields implied by supersymmetry as dictated by\eqref{eq:SusyRelation0}.  The metric can be written in the form
$$
\dd s^2 = 2 G_{\xi\eb}\, \dd y^\xi \dd y^\eb + 2G^0_{\a\bb}\, \dd z^\a \dd z^\bb~,
$$
 where
 \beq
\begin{split}
G^0_{\a\bar\b} ~&=~
- \frac{\displaystyle \int \chi_\a {\bar\chi}_{\bar\b}} {\displaystyle \int \O\, \overline \O}~,\\[10pt]
 G_{\xi\eb}  ~&=~ \frac{1}{V} \int \ccD_\xi \o\, \star\, \ccD_\eb \o +\frac{\ii \ap}{8V}\int \o^2 \tr \Big( \ccD_\xi \A\, \ccD_\eb \A^\dag\Big) - \frac{\ii \ap}{8V}\int \o^2 \tr \Big(\ccD_\xi \vth\, \ccD_\eb \vth^\dag\Big)~.
\end{split}\label{eq:IntroMetric1}
\eeq
Here the $\chi_\a$ form a basis of closed $(2,1)$-forms, and the second term in the last line is the Kobayashi metric, extended to the entire parameter space. The metric is the natural inner product of $\ccD_\x \o$ and $\ccD_\x \A$ together with the inner product of representatives of deformations of complex structure. As expected, the B-field does not make an explicit appearance, being determined by the other fields in the heterotic structure through the anomaly and supersymmetry constraints. The construction of this metric did not assume any underlying special geometry, and its simplicity leads us to conjecture it holds for a general heterotic structure satisfying the equations of motion and so for the Strominger system. The result \eqref{eq:IntroMetric1} builds on \cite{Anguelova:2010ed} who studied the moduli space metric to $\cO(\ap^2)$ restricted to a locus of the parameter space in which only the hermitian part of the metric varies $\d g_{\m\nb} \ne 0$ with the remaining fields remain fixed, $\d A = \d B = 0$. On this sub-locus, the leading correction to the moduli space metric is $\cO(\ap^2)$ and not $\cO(\ap)$. As can be seen from \eqref{eq:IntroMetric1}, this result is a manifestation of demanding the gauge field remain fixed --- which is in our language $\ccD_\x B = 0$ and $\ccD_\x A = 0$. In general, we need to allow all the fields to vary, even when considering  \K parameter variations. As shown in \eqref{eq:IntroMetric1}, this means the metric is corrected at $\cO(\ap)$, with the property, for example, that the special geometry metric is corrected through a mixing the complex structure and hermitian parameter sectors. 

Although the form of this metric is largely determined by demanding gauge invariance, we check  it is also the metric that arises in string theory by dimensionally reducing the $\ap$-corrected supergravity action. For this, we assume $\ccX$ is a \cym, with Ricci-flat metric at zeroth order in $\ap$. Again, the use of supersymmetry is crucial in the reduction, to next order in $\ap$, and it is only after taking it carefully into account does one arrive at the \K metric \eqref{eq:IntroMetric1}. We conjecture this result holds to higher orders in $\ap$, once the variations of the spin connection are included appropriately. As the effective action is known at order $\ap^2$, this can be checked. A naive guess of the form of the metric might be that the metric would have the form of the Kobayashi metric fibred over a special geometry base upon which the Atiyah constraint is imposed. That the metric \eqref{eq:IntroMetric1} does not have this form is a consequence of the fact that, owing to the anomaly conditions, the parameter space does not simply correspond to
$H^1(\ccT_\ccX)\oplus H^1(\ccT^*_\ccX)\oplus H^1(\text{End}\ccT_\ccX)$.

Several heterotic vacua which give the standard model in the four-dimensional world  have been constructed by traditional means. For explicit models, see for example \cite{Anderson:2013xka} and references therein, which at low-energies include the standard model of particle physics. The metric we wish to describe here plays a central role in analysing these constructions. It defines the kinetic energy terms for the massless neutral scalar fields coming from the moduli of the vacuum, and normalises the Yukawa couplings for the charged matter fields.  Both ingredients are essential for a correct description of the four-dimensional theory. It would be interesting to see how the results here influence these constructions. 

The layout of this paper is as follows.
In \SS2 we review the formalism of holomorphic vector bundles. We then turn, in \SS3, to the definition of gauge covariant derivatives on the moduli space. These allow us to write the metric in a manifestly gauge covariant form. In \SS4 we establish relations between the first and second variations of the B-field, gauge field and hermitian form as dictated by supersymmetry. In \SS5, by positing the \K potential \eqref{eq:KahlerPot0} we derive the metric \eqref{eq:IntroMetric1}. This checks the \K potential. Finally, in \SS6, we perform the dimensional reduction of heterotic supergravity, and establish that the metric \eqref{eq:IntroMetric1} is indeed the correct \K metric on the heterotic moduli space. 
\subsection{Notation}
We give here, for reference, two tables: a listing of the coordinates and indices for the spaces that arise, and a listing of notation for the principal quantities of interest. 
\def\medstr{\vrule height19pt depth11pt width0pt}
\def\smallstr{\vrule height16pt depth8pt width0pt}
\begin{table}[H]
\begin{center}
\begin{tabular}{| >{\hskip2pt} l <{} >{$}l<{$} | >{$\hskip-3pt}c<{\hskip-3pt$} | 
>{$\hskip-3pt}c<{\hskip-3pt$} |}
\hline
\multispan2{\medstr\vrule\hfil Coordinates\hfil\vrule}&~\text{Holomorphic Indices}~&~\text{Real Indices}~\\ 
\hline\hline
\smallstr Bundle                   & w^i                  &   i,\,j,\ldots      & I,\, J\ldots \\
\hline
\smallstr Complex structure & z^\a                  & \a,\,\b\ldots     & a,\,b,\ldots \\
\hline
\smallstr \K class                 & t^\r;~t=u+\ii v & \r,\,\s,\ldots      & r,\,s,\ldots \\
\hline
\smallstr Generic holomorphic parameters   & y^\x = (w^i, z^\a, t^\r)   &\x,\,\eta, \ldots      & M,\,N,\ldots \\
\hline
\smallstr \cym                     & x^\m                & \m,\,\n,\ldots   &m,\, n,\ldots \\
\hline
\smallstr Minkowski space    & X^e                   &                       &e,\, f,\ldots \\
\hline
\end{tabular}
\capt{5.0in}{tab:coords}{The coordinates and indices for the various spaces.}
\end{center}
\end{table}
\vfill
\begin{table}[H]
\begin{center}
\begin{tabular}{| >{$\hskip1pt} l <{\hskip1pt$} | >{\hskip1pt}l<{\hskip1pt} 
| >{\hskip1pt}l<{\hskip1pt} | l |}
\hline
\medstr\hfil $ Quantity $ & \hfil Comment & \hfil Definition & \hfil Ref.\\ 
\hline\hline
\smallstr  A~=~\cA - \cA^\dag         &   $\cA$ is the $(0,1)$-part    & Yang-Mills gauge potential &\\
\hline
\smallstr  \Th~=~\vth - \vth^\dag   &   $\vth$ is the $(0,1)$-part    & Lorentz gauge potential    &\\
\hline
\smallstr  F,\, R                               &                                            & Field strength for $A$ and $\Th$ &\\
\hline
\smallstr  \Ph,\,\Ps                          &                                            & YM and Lorentz gauge functions  &\\
\hline
\smallstr  Y,\,Z                          & & $Y=\Ph^{-1}\dd\Ph,~Z=\Ps^{-1}\dd\Ps$ &\sref{sec:gaugeconn}\\
\hline
\smallstr  Y_I,\,Z_I                   & & $Y_I=\Ph^{-1}\del_I\Ph,~Z_I=\Ps^{-1}\del_I\Ps$ &\sref{sec:derivhet}\\
\hline
\smallstr  U,\,W                              &  two forms       
                & $\dd U=\smallfrac13\tr(Y^3)$ ~and~ $\dd W=\smallfrac13 \tr(Z^3)$   &\sref{sec:BHtransfs}\\
\hline
\smallstr  U_M,\,U_{MN}                &  two forms       
                & related to the covariant derivatives of $B$   &\sref{sec:derivB}\\                
\hline
\smallstr  \L~=~\l - \l^\dag  & $\l$ is the $(0,1)$-part 
                & $y^M$-components of the YM potential  &\sref{sec:derivhet}\\
\hline
\smallstr  \P                                   &                           
                & $y^M$-components of the Lorentz potential  &\sref{sec:derivhet}\\
\hline
\smallstr  \IA,\, \ITheta   &  & $\IA=A+\L,~\ITheta=\Th+\P$ &\sref{sec:params},\sref{sec:UnivBundle}\\                
\hline
\smallstr  \IF,\, \IG          &  & Field strengths of $\IA$ and $\ITheta$ &\sref{sec:derivhet}\\
\hline
\smallstr  \dd_A\! =\,\delb_{\!\cA} - \del_{\!\cA^\dag} 
                &  $\delb_\cA$ is the $(0,1)$-part & Gauge covariant exterior derivative  &\sref{sec:gaugeconn}\\
\hline
\smallstr D_\x,~D_I                  &   & Special geom.\ and gauge cov.\ derivatives &\sref{sec:params} \\
\hline
\smallstr  \ccD                          &                  & Holotypical derivative                         &\sref{sec:holoD} \\
\hline
\smallstr  b_M                          & one form    & relates to the gauge transform of $B$    &\sref{sec:derivB} \\
\hline
\smallstr  \ccB_\x                     & two form    & relates to the holotypical derivative of $B$ &\sref{sec:derivB} \\
\hline
\end{tabular}
\capt{5.0in}{tab:notation}{A table of notation for the principal quantities that we consider.}
\end{center}
\end{table}
\newpage
\section{Holomorphic vector bundles}
The following is a review of the general formalism associated with holomorphic vector bundles whose purpose is largely to set out our conventions and notation.
\subsection{Antihermitean gauge connections}\label{sec:gaugeconn}
Let $\ccE$ denote a vector bundle, with structure group $\eu{G}$, over a manifold $\ccX$ and let~$A$ be the corresponding gauge potential.  So $A$ is a one--form valued in the adjoint representation of the Lie algebra of $\eu{G}$.
Under a gauge transformation, $A$ has the transformation rule
\beq
A \to \Ph A \Ph^{-1} -\, \dd\Ph\,\Ph^{-1}
\label{GaugeTransf}\eeq
where $\Ph$ is a function on $\ccX$ that takes values in $\eu{G}$. If we take $\Ph$ to be unitary then $ \dd\Ph\,\Ph^{-1}$ is antihermitean and so, with the transformation rule as above, is $A$. We could redefine $A$ by a factor of $\ii$, so as to make it hermitean, but this would be at the expense of introducing corresponding factors into later expressions.

Under a gauge transformation, a zero-form $\ps$, that transforms in the fundamental representation of the gauge group, transforms as $\ps\to\Ph\ps$. The gauge potential serves to define a covariant derivative
$$
\dd_A\ps~=~(\dd + A)\ps~,
$$
which transforms like $\ps$, that is $\dd_A\ps \to \Ph\, \dd_A\ps$.
The field strength of $A$ is the curvature of the connection
$$
\dd_A^2\ps~=~F\, \ps~,
$$
where
$$
F~=~\dd A + A^2~.
$$
The field strength transforms in the adjoint of the gauge group: $F\to \Ph F\Ph^{-1}$.

Let $\A$ be the $(0,1)$ part of $A$ then, since $A$ is antihermitean,
$$
A = \A - \A^{\dag}~.
$$
On decomposing the field strength into type. We find
\beq\begin{split}
F^{(2,0)}~&=-\del \A^\dag + (\A^\dag)^2\\[3pt]
F^{(1,1)}~&=~\del\A - \delb\A^\dag - \{\A, \A^\dag\}\\[3pt]
F^{(0,2)}~&=~\delb\A \;+\, \A^2~.
\end{split}\label{Fexpr}\eeq
The bundle $\ccE$ is holomorphic if and only if there exists a connection such that $F^{(0,2)}=0$. The Hermitean Yang-Mills equation is
$$
\o^2 F ~=~0~.
$$

The definition of the covariant derivative is extended so as to conform with the Leibnitz rule, and the precise form of the derivative then depends on the transformation properties of the object on which it acts. In the following we will want the derivative to act on $p$-forms that take values in $\End(\ccE)$. Denoting such a quantity by 
$\b_p$, the derivative takes the form
$$
\dd_A\, \b_p = \dd \b_p + A\, \b_p - (-1)^p\, \b_p\, A~.
$$
We denote by $\delb_{\!\cA}$ and $\del_{\!\cA^\dag}$ the $(0,1)$ and $(1,0)$ parts of $\dd_A$. Thus
$$
\delb_{\!\cA}\, \b_p~=~\delb \b_p + \cA\, \b_p - (-1)^p\, \b_p\, \cA~,~~~\text{and}~~~
\del_{\!\cA^\dag}\,\b_p~=~\del\b_p - \cA^\dag\b_p + (-1)^p \b_p\, \cA^\dag~.
$$
The connection $A$ has a gauge transformation property
\beq
^{\Phi}A = \Phi (A - Y)\Phi^{-1}~, \quad Y=\Phi^{-1}\dd \Phi~,
\notag\eeq
where $\Phi$ is a gauge function that depends on the coordinates of $\ccX$ and the parameters defining the heterotic structure. We demand that all physical quantities be covariant under this transformation, and in particular, the field strength $F$. 

Consider a deformation $A\to A{+}\d A$, together with a variation of the gauge transformation 
$\Phi\to\Phi(1+\e)$. 
The total quantity $A+\d A$ transforms by a rule analogous to \eqref{GaugeTransf}. We are free, however, to distribute the `blame' for the variation of the gauge transformation between $A$ and $\d A$. We follow the doctrine of the background field method in assigning the blame entirely to $\d A$. In this way we find that $A$ continues to transform according to \eqref{GaugeTransf}, while
$$
\d A~\to~ \Ph\big( \d A - \dd_A\e\big) \Ph^{-1}~.
$$
This  is understood as the  composition of two gauge transformations:
\begin{itemize}
\item Background gauge transformations
$$
A~\to~ \Ph A \Ph^{-1} -\, \dd\Ph\,\Ph^{-1}~,~~~\text{and}~~~
\d A~\to~ \Ph\, \d A\, \Ph^{-1}~.
$$

\item Small gauge transformations
$$
A~\to~A~,~~~\text{and}~~~\d A~\to~\d A -\dd_A\e~.
$$
\end{itemize}
For the remainder of the paper we will define $\d A$ such that it is gauge covariant fashion transforming  according to background gauge transformation law. We will define this more exactly in the next section. 

Consider now a holomorphic bundle over a fixed $\ccX$, that is for which $F^{0,2}=0$. Then, decomposing $\d A$ into type $\d A=\d\cA-\d\cA^\dag$, demanding that $F^{0,2}=0$ be preserved leads to 
$$
\delbA\, \d\cA~=~0~,
$$
where
$$
\delbA\, \d\cA~=~\delb\, \d\cA + \{\cA,\, \d\cA\}
$$
is the $(0,2)$ part of $\dd_A\, \d\cA$. Under a small gauge transformation, where $\Phi \approx 1+ \e$, we have $\d\cA\to\d\cA + \delbA\e$ and we learn that the first order deformations of $\cA$ correspond to the cohomology of $\delbA$. 

We would like to understand the relation of this cohomology to that of $\delb$ and to understand also how the condition $F^{(0,2)}=0$ is equivalent to the fact that the bundle is holomorphic. We will proceed, as above, in terms of locally defined quantities.
\subsection{The gauge prepotential}
Locally, the equation $\delb\cA + \cA^2 = 0$ is solved by taking
\beq
\cA~=-\delb \m\,\m^{-1}~,
\label{GaugePrepotential}\eeq
for some matrix $\m$ valued in the Lie algebra of $\eu{G}$. The matrix $\m$ need not be unitary so
\beq
\cA^\dag~=-(\m^\dag)^{-1}\del \m^\dag~,
\label{GaugePrepotentialDag}\eeq
and it is only when $\m$ is unitary that $A$ is pure gauge.

The relation between the cohomology groups of $\delb_{\!\cA}$ and $\delb$ follows from the observation that, given \eqref{GaugePrepotential},
$$
\m^{-1}\delb_{\!\cA}\, (\d\cA)\,\m~=~\delb(\m^{-1} \d\cA\, \m)~~~\text{and}~~~\m^{-1}\delb_{\!\cA} \e\,\m~=~\delb(\m^{-1} \e \m)~.
$$
These relations provide the isomorphism between the cohomology of $\delb_{\!\cA}$ and that of $\delb$.

The quantity $\m$  can be chosen such that, under a gauge transformation, it transforms according to the rule
\beq
\m~\to~\Ph \m~.
\label{UnitaryGaugeTransf}\eeq
The matrix $\m$ is not uniquely determined by the gauge potential since $\m$ and $\m\z$, with $\z$ a holomorphic matrix, determine the same $\cA$.
We shall refer to the replacement
\beq
\m~\to~\m \z
\label{HolGaugeTransf}\eeq
as a holomorphic gauge transformation.

\subsection{Complex gauge transformations}
In applications to string theory vacua it is natural to take the group $\eu{G}$ to be a compact Lie group, and so a group represented by unitary matrices. The gauge potential is then antihermitean. The theory of holomorphic vector bundles can however be developed for complex groups and there is then no need to take the gauge potential to be antihermitean. So now we have $A=\cA+A^{(1,0)}$, but now $A^{(1,0)}$ is not, in general, 
$-\cA^\dag$. It is common, in this context, to adopt a different, but equivalent, definition for a holomorphic bundle, as a vector bundle for which $\cA = A^{(0,1)}$ vanishes. Let us write an antihermitean gauge potential $A=\cA-\cA^\dag$ in terms of the gauge prepotential, via \eqref{GaugePrepotential} and \eqref{GaugePrepotentialDag}, and consider the effect of a general (nonunitary) gauge transformation, of the form~\eqref{GaugeTransf}
\beq\begin{split}
A~\to~{}^\Ps\!A~&=\Ps\big(-\delb\m\,\m^{-1} + (\m^\dag)^{-1}\del\m\big)\Ps^{-1} - \dd\Ps \Ps^{-1}
\\[5pt]
&=-\delb(\Ps \m)\,(\Ps \m)^{-1} + (\m^\dag \Ps^{-1})^{-1}\,\del (\m^\dag \Ps^{-1})~.
\end{split}\label{GenGaugeTransf}\eeq
We achieve a holomorphic frame by taking $\Ps=\m^{-1}$, the more general solution $\Ps=(\m\z)^{-1}$, with   $\z$ holomorphic, is equivalent to first taking $\Ps=\m^{-1}$ and then making a holomorphic gauge transformation. The gauge potential is then given by
$$
A_\text{hol}~=~(\m^\dag \m)^{-1}\,\del (\m^\dag \m)~.
$$
Note that $A_\text{hol}$ is invariant under the unitary gauge transformation \eqref{UnitaryGaugeTransf} but that under the holomorphic gauge transformation \eqref{HolGaugeTransf} we have
$$
A_\text{hol}~\to~ \z^{-1}A_\text{hol}\, \z - \del( \z^{-1})\, \z~,
$$
which justifies the terminology. The holomorphic gauge transformations are precisely the gauge transformations that preserve the holomorphic frame.

Just as we can choose a holomorphic frame with $\cA=0$, so we may also choose, what we can refer to as antiholomorphic frame, with $A^{(1,0)}=0$. To achieve this gauge, we take $\Ps=\m^\dag$ in \eqref{GenGaugeTransf}. This gives
$$
A_{\overline{\text{hol}}}~=-\delb (\m^\dag \m)\, (\m^\dag \m)^{-1}~.
$$
This gauge is preserved by the antiholomorphic gauge transformations
$$
A_{\overline{\text{hol}}}~\to~\z^\dag A_{\overline{\text{hol}}}\, (\z^\dag)^{-1} -
\delb\z^\dag\, (\z^\dag)^{-1}~.
$$

To simplify notation we will use $\ccA$ and $\ccm$ to denote the gauge potential and the gauge prepotential in antiholomorphic frame. Thus
$$
A_{\overline{\text{hol}}}~=~\cA_{\overline{\text{hol}}}~=~\ccA~~~\text{and}~~~
\m_{\overline{\text{hol}}}~=~\m^\dag\m~=~\ccm~.
$$
We take $\m$ to have the transformation law
$$
\m~\to~\Ph\m\z~,
$$
with $\Ph$ unitary and $\z$ holomorphic. Thus, under a gauge transformation, 
$$
\ccm~\to~\z^\dag \ccm\,\z
$$
and the transformation rules for $\cA$ and $\ccA$ follow.

In the antiholomorphic frame we have
$$
F_{\overline{\text{hol}}}~=~\del\ccA
$$
so the condition $g^{\m\bar\n}F_{\m\bar\n}=0$ is equivalent to
$$
g^{\m\bar\n}\del_\m(\del_{\bar\n}\ccm\,\ccm^{-1})~=~0~.
$$
Note that, owing to the relation $\ccm=\m^\dag\m$, with $\m$ invertible, the matrix
$\ccm$ is hermitean and positive definite. In the mathematics literature $\ccm$ is often referred to as a `metric', though to do so here would invite confusion.

The Uhlenbeck-Yau theorem states that, for a stable holomorphic bundle, there exists a solution to the equation above that is unique up holomorphic gauge transformations 
$\ccm\to\z^\dag\ccm\z$. Given a positive definite, hermitean matrix $\ccm$, there is a natural choice for the square root $\sqrt{\ccm}$, and the general solution for $\m$ to the equation $\ccm=\m^\dag\m$ is $\m=\Ph\sqrt{\ccm}$, with $\Ph$~unitary.

Finally, let us make some brief observations about the antiholomorphic frame. If we subject $\L$ to a complex gauge transformation with gauge function $\Ph=\m^\dag$, in order to bring it to the antiholomorphic frame, then the result is no longer antihermitean. We find that
$$
\L_j^{\overline{\text{hol}}}~=~0~~~\text{and}~~~
\L_{\bar k}^{\overline{\text{hol}}}~=-\del_{\bar k}\ccm\,\ccm^{-1}
$$
It is immediate that
$$
D_j\ccA~=~\del_j\ccA~~~\text{and}~~~D_{\bar k}\ccA~=~0~.
$$
\subsection{Principal bundles}
Closely associated with gauge connections for a vector bundle is the notion of a principle bundle. This is a bundle for which the fibre is a Lie group $\eu{G}$. The points of the fibre correspond to elements $h\in \eu{G}$. Gauge transformations will be associated with translations $h\to \Ph h$ in the group.

The connection one-form is defined as
$$
\S~=~h^{-1} \dd h + h^{-1}A\,h~.
$$
Note that $\S$ is invariant under a gauge transformation provided $A$ transforms as in \eqref{GaugeTransf}. The curvature two-form is
$$
\dd\S + \S^2~=~h^{-1} F\, h~,
$$
which is also invariant.

We decompose the connection one form into its $(0,1)$ and $(1,0)$ parts $\S = \s - \s^\dag$ with
$$
\s~=~h^{-1}\delb h + h^{-1}\!\cA\,h~~~\text{and}~~~
\s^\dag~=~-h^{-1}\del h + h^{-1}\!\cA^\dag h~.
$$
On the principal bundle the natural background gauge invariant metric is
$$
\dd s^2~=~2G_{\m\bar\n}\, \dd x^\m\otimes\dd x^{\bar\n} +
\frac{\ap}{4}\tr\left( \s\otimes\s^\dag\right)~.
$$

\subsection{Complex structure of the bundle}
We take the bundle to vary holomorphically with parameters $w^j$. This being so, it is natural to allow the gauge potential $A$, and the gauge function $\Phi$ to depend on the parameters. We will take $\A$ to depend holomorphically on parameters in the following sense. The operator $\delb_\A$ defines a complex structure on the bundle and the vanishing of $\delb_{\!\cA}^2$ is then the condition that this complex structure be integrable. In the mathematics literature, one often chooses a holomorphic gauge in which $\delb_\A$ reduces to $\delb$. This is expanded upon in the appendix, but do not make this choice here. A section $s$ of the bundle is said to be holomorphic if the covariant derivative $\delb_{\!\cA} s$ vanishes. Such a section will not be holomorphic in the naive sense that $\delb s=0$. Indeed, such a condition is not invariant under gauge transformation. In a similar way, the fact that the bundle can be taken to depend holomorphically on parameters $w^j$ does not mean that $\partial_{\bar\jmath}\cA=0$, since, as we shall see, this condition is not invariant under gauge transformation. Rather the condition is $\ccD_{\bar\jmath}\cA=0$ where $\ccD_{\bar\jmath}$ denotes a suitably defined covariant derivative with respect to the parameters. It is to the definition of this covariant derivative that we turn in \sref{sec:params}.

\subsection{The $B$ and $H$ fields\label{sec:BHtransfs}}
The heterotic geometry defines a gauge-invariant three-form
\beq
H~=~\dd B - \frac{\ap}{4}\Big(\CS[A] - \CS[\Th]\Big)~,
\label{Hdef}\eeq
where $\CS$ denotes the Chern-Simons three-form
$$
\CS[A]~=~\tr\!\left(A\dd A +\smallfrac23\, A^3\right)~=~\tr\!\left(AF - \smallfrac13\, A^3\right)~,
$$
and $\Th$ is the gauge potential for frame transformations. The three-form $\dd B$ is defined so that  $H$ is gauge invariant, and so $\dd B$ itself has gauge transformations and is viewed as being slaved to the geometry of the $\ccX$ and the gauge bundle.  

Under background gauge and Lorentz transformations, we have 
$$
A~\to~\Ph (A - Y)\Ph^{-1}~~~\text{and}~~~F~\to~\Ph F \Ph^{-1}~;
~~~\text{with}~~~Y~=~\Ph^{-1}\dd\Ph~,
$$
and similarly
$$
\Th~\to~\Ps (\Th - Z)\Ps^{-1}~~~\text{and}~~~R~\to~\Ps R \Ps^{-1}~;
~~~\text{with}~~~Z~=~\Ps^{-1}\dd\Ps~.
$$
Noting the identities $\dd Y=-Y^2$ and $\dd Z=-Z^2$, we see that
$$
\CS[A]~\to~\CS[A] - \dd\tr\!(AY) + \frac13 \tr\left(Y^3\right)~,
$$
together with the analogous rule for $\CS[\Th]$.

Now the integral of $\tr(Y^3)$ over a three-cycle is a winding number, so vanishes if the gauge transformation is continuously connected to the identity. Since the integral vanishes for every three-cycle we have that 
$\tr(Y^3)$ is exact
$$
\frac13\,\tr(Y^3) ~=~\dd U~,~~\text{and similarly:}~~~\frac13\,\tr(Z^3) ~=~\dd W~,
$$
for some globally defined two forms $U$ and $W$.

The anomaly cancellation condition means that the $B$ field is assigned a transformation so as to cancel the derivative terms that arise from the Chern-Simons forms
$$
B~\to~B - \frac{\ap}{4}\left\{\tr(AY) - U - \tr(\Th Z) + W\right\}~.
$$
With this transformation law, $H$ is invariant.
\vskip15pt
\begin{table}[H]
\def\bigstr{\vrule height22pt depth12pt width0pt}
\def\medstr{\vrule height18pt depth10pt width0pt}
\def\smallstr{\vrule height17pt depth10pt width0pt}
\begin{center}
\begin{tabular}{| >{~$} l <{$~} | >{~$}l<{$~} |}
\hline
\multispan2{\medstr\vrule\hfil\large Gauge transformations\hfil\vrule}\\
\hline
\noalign{\vskip10pt}
\hline
\multispan2{\smallstr\vrule\hskip10pt Background gauge transformations\hfil\vrule}\\ 
\hline\hline
\smallstr A~\to~\Ph\,(A-Y)\,\Ph^{-1}                     & \d A~\to~\Ph\, \d A\, \Ph^{-1} \\
\hline
\smallstr \Th~\to~\Ps\,(\Th-Z)\,\Ps^{-1}               & \d\Th~\to~\Ps\, \d\Th\, \Ps^{-1} \\
\hline
\smallstr B~\to~B - \frac{\displaystyle\ap}{\displaystyle 4}\Big(\!\tr \big(AY - \Th Z\big) - U + W\Big)  & 
\d B~\to~\d B -\frac{\displaystyle\ap}{\displaystyle 4}\tr\big( \d A\, Y - \d\Th\,Z \big) \\
\hline
\noalign{\vskip10pt}
\hline
\multispan2{\smallstr\vrule\hskip10pt Small gauge transformations\hfil\vrule}\\ 
\hline\hline
\smallstr A~\to~A                                                   & \d A~\to~\d A - \dd_A\e \\
\hline
\smallstr \Th~\to~\Th                                             & \d \Th~\to~\d \Th - \dd_\Th\eta \\
\hline
\smallstr B~\to~B                                                   & \d B~\to~\d B - 
\frac{\displaystyle\ap}{\displaystyle 4}\tr\big( A\, \dd\e - \Th\, \dd\eta\big) \\
\hline
\end{tabular}
\capt{4.5in}{gaugetransformations}{The transformation rules for the gauge potentials $A$ and $\Th$, and 
the $B$-field under background and small gauge transformations.}
\end{center}
\end{table}
\newpage
\section{Rudiments of the parameter space geometry}\label{sec:params}
Deformations of the gauge field, $A\to A+ \d A$, have two sources: those which are related to the parameters of  $\ccX$, and those which cannot be undone by changing the parameters of the manifold. The former derive from the complex structure moduli $z^\a$ and the \K moduli $t^\r$ of $\ccX$. The latter are the bundle moduli, $w^i$, and are related to endomorphisms of the bundle $\ccE$. 

The deformations must preserve the equations of motion, and so  the bundle must remain holomorphic and solve the HYM equation.  Holomorphy  is a closed condition in complex structure moduli space; that is, there are directions in the parameter space which are not allowed. Hence, only a subset of the complex structure moduli of $\ccX$ are actual parameters of the heterotic compactification. The satisfaction of the HYM equation is open in complex structure moduli space, and does not restrict any of the complex structure moduli. We always assume we are deforming about a supersymmetric solution  of the equations of motion  away from the stability walls discussed in  \cite{Anderson:2009nt} say.  Hence, the HYM equation does not obstruct any of the \K moduli. In other words, $F$ obstructs some of the complex structure moduli $z^\a$, a phenomenon well-known from the type II flux compactification literature \cite{Grana:2005jc}, while the \K parameters  $t^\r$ and bundle moduli $w^i$ are unobstructed. 
Since we allow the parameters of the triple $(\ccX, \ccE, H)$ to vary it is natural to allow gauge transformations that depend on these parameters. Doing so means we should introduce a connection $\L$,  a 1-form on the parameter space $\cM$, in order to define appropriate covariant derivatives with respect to the parameters. 
It is then natural to unify $\L$ and $A$ by defining a connection $\IA = A {+} \L$, which serves as a connection for a fibre bundle whose base space can be locally written as $\ccX{\times}\cM$. This bundle, known as a universal bundle, was first introduced in the Atiyah-Singer index theorem \cite{AtiyahSinger} and in the physics literature through the study of BPS monopoles, for example \cite{Gauntlett:1993sh}. We outline the consequences of this identification in \sref{sec:UnivBundle} and it is fully explored in upcoming work \cite{CdOMcO:2016a}.  In our case we want the total space $(\ccX,\ccE)$ together with $H$ to vary with parameters. We call the triple $(\ccX,\ccE,H)$ a heterotic structure so that the universal bundle is a family of heterotic structures $(\ccX, \ccE, H)_{(z,w,t)}$ over $\cM$. This means that for each $(z,w,t)\in \cM$, there is a corresponding vector bundle $\ccE$ with base 
$\ccX$.

This situation is already  familiar from the study of the complex structure moduli of CY manifolds. As described by Kodaira--Spencer, one has a family of CY manifolds $M_z$  over a moduli space of complex structures $\cM_z$, and these form a complex analytic family of complex manifolds. The holomorphic three-form $\O$ is a section of a line bundle over $\cM_z$, and so variations of $\O$ are described by a covariant derivative of $\O$.  The covariant derivative of $\O$ is similar to the covariant derivatives of the connection $A$ we describe here. 

We will largely be non-specific about the division of labour between the parameters $(z,w,t)$. In fact, where possible our results will be stated for a general parameter $y{\,=\,}(z,w,t)$, reflecting the fact that there is generally no unambigious separation between these parameters.

\subsection{A review of  special geometry}\label{SpecialGeom}
We wish to develop heterotic geometry, the geometry of the moduli space of heterotic vacua.  In order to do this we shall need to construct the moduli space metric and certain covariant derivatives. This geometry will then be a generalisation of special geometry. We find it useful, therefore, to briefly review special geometry before proceeding to the more general~case. 

The parameter space reduces to that of special geometry when the bundle $\ccE$ is identified with the tangent bundle. The parameter space splits into those corresponding to complex structure deformations and those corresponding to \K deformations. The metric on the space of these two types of deformations takes the form 
$$
\dd s^2 = 2G^0_{\a\bb}\, \dd z^\a \dd z^\bb + 2G^0_{\r\sb}\, \dd t^\r \dd t^\sb,
$$  
where  $G^0_{\a\bar\b}$ and $G^0_{\r\bar\s}$ are \K metrics whose \K potentials are
\beq
K_\CS~=-\log\left(-\ii \int\! \O\,\overline{\O}\right)~~~\text{and}~~~ 
K_\text{K\"ah}~=-\log\left(\frac43\int\! \o^3\right)~,
\label{Kpotentials}\eeq
respectively. In these expressions $\O$ is the holomorphic (3,0)-form and $\o$ denotes the \K-form. 
These metrics arise naturally in two ways. The first is to write down the natural metric on the space of Ricci-flat metrics, augmented to include also the variations of the $B$-field. 
\beq
\dd s^2 = \frac{1}{4V}\int \dd^6x\, \sqrt{g} \, g^{km} g^{ln}
\big(\d g_{kl}\,\d g_{mn} + \d B_{kl}\, \d B_{mn}\big)~,
\label{deWitt}\eeq
Here $g_{mn}$ and $g_{mn}+\d g_{mn}$ are Ricci-flat metrics, and the variations 
$\d g_{mn}$ and 
$\d B_{mn}$ are subject to the constraints $\nabla^m \d g_{mn}=0$ and 
$\nabla^m \d B_{mn}=0$. All quantities here receive $\ap$ corrections; these corrections will be crucial in studying the heterotic moduli space. 

In virtue of Yau's theorem, the Ricci-flat metrics on the real manifold $\ccX$ are in one-one correspondence with the members of the family of complex manifolds $M(z,t)$. Apart from the prefactor of $1/4V$ and the $\d B$-terms, this metric seems to have been first written down by deWitt in his early considerations of the path integral for quantum gravity. On decomposing the metric variation into complex type,
$\d g_{mn}=\{\d g_{\m\n}, \d g_{\m\bar\n}, \d g_{\bar\m\bar\n}\}$, the deWitt metric separates into a metric on the complex structure parameters, the part corresponding to the pure parts $\d g_{\m\n}$, 
$\d g_{\bar\m\bar\n}$, and a remainder which corresponds to a metric on the \K-class parameters. 

This same metric on the complex structures can be derived also from the consideration that the holomorphic three form $\O$ determines the complex structure and may be chosen so as to vary holomorphically with the parameters. The scale of $\O$ is undefined so there is a natural `gauge invariance'
\beq
\O~\to~f(z)\,\O~,
\label{OmegaGauge}
\eeq
where $f(z)$ is any holomorphic function of the complex structure parameters. The natural metric constructed from $\O$ that is invariant under these gauge transformations is the \K metric corresponding to the \K potential
$K_\CS$ given in \eqref{Kpotentials}. For a more detailed introduction to special geometry, in the style of the present article, see \cite{Candelas:1990pi}. The \K-parameter part of the deWitt metric \eqref{deWitt} can be shown to be a \K metric corresponding to a \K potential $K_\text{K\"ah}$.

Under a variation of complex structure, the holomorphic three-form varies into a part that is again of type $(3,0)$ and a part that is of type $(2,1)$
\beq
\pd{\O}{z^\a}~=-k_\a\,\O + \ch_\a~;~~~
\chi_\a~=~\frac12 \chi_{\a\,\m\n\bar\r}\,\dd x^\m \dd x^\n \dd x^{\bar\r}~.
\label{Kodaira}
\eeq
The minus sign that precedes $k_\a$ has been chosen to simplify a later relation and the $(2,1)$-forms 
$\chi_\a$ are vectors on the parameter space, corresponding to complex structure variations and appear repeatedly in the following. In particular, the $\chi_\a$ make a prominent appearance in the metric on the space of complex structures. It is straightforward to show that the complex structure part of the deWitt metric may be rewritten in the form
\beq
G^0_{\a\bar\b} ~=~
- \frac{\displaystyle \int\! \chi_\a {\bar\chi}_{\bar\b}} {\displaystyle \int\! \O\, \overline \O}~. 
\label{SGCSMetric}\eeq

The metric on the \K class parameters has, at first sight, a rather different form
$$
G^0_{\r \bar\s} ~=~ \frac{1}{4V} \int\! e_\r\! \star e_{\bar\s}~,\label{SGKahlerMetric}
$$ 
where $e_\r$ $(=e_{\bar\r})$ are a basis for $H^2(\ccX,\IZ)$.

Standard coordinates $z^a,~a=0,1,\ldots,h^{1,2}$, the so-called special coordinates, are obtained for the complex structures by taking a basis of 3-forms dual to a symplectic basis for $H_3(\ccX,\IZ)$ and writing
$$
\O~=~z^a\a_a - \cG_b(z)\,\b^b~.
$$
It may be shown that $\cG_a = \partial\cG/\partial {z^a}$ for a function $\cG(z)$, known as the prepotential, that is homogeneous, of degree two, in the $z^a$. Here the $z^a$ are projective coordinates for the complex structure and our use of the index $a$ is at variance with the use in the rest of this article. For the complexified \K class we form the complex combination $B+\ii\o$ and write
$$
B+\ii\o~=~t^\r\,e_\r~,
$$
where the $e_\r$ form a basis for $H^2(\ccX,\IZ)$. We may define a prepotential for the \K parameters by setting
$$
\cF_0(t)~=-\frac{1}{3!}\,y_{\r\s\t}\frac{t^\r t^\s t^\t}{t^0}~. 
$$
We have added an extra coordinate $t^0$ in order to render $\cF_0(t)$ also homogenous of degree two. The suffix 0 on the \K prepotential reminds us that the prepotential receives quantum corrections so that the quantum corrected prepotential is a deformation of $\cF_0$, so is of the form $\cF=\cF_0 + \D\cF$. The prepotentials determine the \K potentials. We have the relations
$$
\ee^{-K_\text{cs}}~=~\ii\left( \bar{z}^a \pd{\cG(z)}{z^a} - z^a \pd{\overline{\cG(z)}}{\bar{z}^a}\right)
~~~\text{and}~~~
\ee^{-K_\text{K\"ah}}~=~\ii\left( \bar{t}^r \pd{\cF_0(t)}{t^r} - t^r \pd{\overline{\cF_0(t)}}{\bar{t}^r}\right)~.
$$
The quantum corrected geometry of the space of \K-parameters is obtained by replacing $\cF_0$, in the above expressions, by the corrected prepotential $\cF$. The fact that the \K potentials above are given by identical expressions in terms of the prepotentials was a surprise when it was first found and was a strong indication of the existence of mirror symmetry. 

An important, but often overlooked, point is that under a complex structure deformation the \K form and B-fields acquire a $(0,2)$-component. For a \cym, the equations of motion for these deformations, together with a suitable choice of gauge fixing, imply these deformations are both exact, and and co-closed. On a compact manifold with holonomy $SU(3)$, rather than a subgroup, such deformations must vanish. This is discussed in more detail in  \sref{eq:DimRedSG}. In heterotic geometry, where the zeroth order geometry is a \cym, we have a nonzero $H$ that arises at $\cO(\ap)$. So we find the $(0,2)$-components of the deformations do not necessarily vanish, though they are at least of order $\ap$. 

\subsection{Covariant derivatives for special geometry}\label{sec:derivSG}
\label{sec:covDeriv}
In the study of complex structures it is important that $\O$ is a section of a line bundle over the moduli space of complex structures, so that it has the \K gauge freedom \eqref{OmegaGauge}. 
We may rewrite \eqref{Kodaira} in the form 
$$
\chi_\a = \del_\a \O + k_\a \O~,
$$
this form suggests that $\chi_\a$ is a covariant derivative of $\O$ with connection $k_\a$.  Indeed, under 
$\O \to f(z)\O$ we have $k_\a$ transforms inhomogeneously, $k_\a \to k_\a + f^{-1} \del_\a f$,  while 
$\chi_\a$ transforms homogeneously, $\chi_\a \to f \chi_a$. By multiplying 
\eqref{Kodaira} by $\overline{\O}$ and integrating over $\ccX$ we find that
\beq
k_\a ~=~\del_\a K^\cs.
\label{CSPotential}
\eeq
The quantity
\beq
D_\a \O ~=~ \del_\a \O + K_{,\a} \O~.
\label{CSCovDeriv}\eeq
is indeed a covariant derivative since it transforms homogeneously, $D_\a\O\to fD_\a\O$, under the \K gauge transformation $\O\to f\O$. 

A quantity $\X^{(a,b)}$ which transforms according to the rule 
\beq
\X^{(a,b)}~\to~f^a \bar{f}^b\, \X^{(a,b)}
\notag\eeq
is said to transform with weight $(a,b)$. For such a quantity the covariant derivative takes the form
\beq\begin{split}
D_\a\X^{(a,b)}~&=~\del_\a \X^{(a,b)} + a K_{,\a}\,\X^{(a,b)}\\[3pt]
D_{\bar\b}\X^{(a,b)}~&=~\del_{\bar\b} \X^{(a,b)} + b K_{,\bar\b}\,\X^{(a,b)}~.
\end{split}\notag\eeq
Note that, in virtue of \eqref{Kpotentials}, the quantity $\ee^{-K}$ has weight $(1,1)$ and $\ee^K$ has weight $(-1,-1)$ with the consequence that
\begin{alignat*}{2}
D_\a\ee^{-K}&=~0~,\quad & D_{\bar\b\,}\ee^{-K}&=~0~,\\[3pt]
D_\a\ee^K~&=~0~,   & D_{\bar\b\,}\ee^K~&=~0~.
\end{alignat*}
A priori the $D_\a \O$ could reside in $H^{3,0}{\oplus}H^{2,1}$ but, in fact, they reside purely in $H^{2,1}$ and moreover form a basis for this space. The derivatives $D_\a \ch_\b$ could, a priori, reside in
$H^{2,1}{\oplus}H^{1,2}$, but these reside purely in $H^{1,2}$. Finally, the derivatives 
$D_\a \ch_{\bar\b}$ could, a priori, reside in $H^{1,2}{\oplus}H^{0,3}$ but reside, in fact, purely in 
$H^{0,3}$. It is straightforward to show that the following relations hold. These can be taken to characterise special geometry.
\beqnn\begin{split}
D_\a\O\hskip4pt~&=~\ch_\a\\[3pt]
D_\a\ch_\b~&=-\ii\,\ee^K y_{\a\b}{}^{\bar\g}\ch_{\bar\g}\\[3pt]
D_\a\ch_{\bar\g}~&=~G^0_{\a\bar\g}\,\overline{\O}\\[5pt]
D_\a\overline{\O}\hskip4pt~&=~0~.
\end{split}\eeqnn
The set $\{\O,\ch_\a,\ch_{\bar\b},\overline{\O}\}$ spans $H^3(\ccX,\IC)$. There is a related `normalised' basis $\{\O,\ch_\a,\widetilde{\ch}^\b,\widetilde{\O}\}$ where
\beqnn
\widetilde{\ch}^\b~=-\ii \,\ee^K \ch^{\b}~~~\text{and}~~~
\widetilde{\O}~=-\ii\,\ee^K\,\overline{\O}~.
\eeqnn
In this basis the nonzero inner products are
\beqnn
\int\!\ch_\a\, \widetilde{\ch}^\b ~=~ \d_\a{}^\b~~~\text{and}~~~\int\! \O\,\widetilde{\O}~=-1~,
\eeqnn
and the special geometry relations assume a somewhat simpler form
\beqnn\begin{split}
D_\a\O\hskip4pt~&=~\ch_\a\\[3pt]
D_\a\ch_\b~&=~y_{\a\b\g}\,\widetilde{\ch}^\g\\[3pt]
D_\a\widetilde{\ch}^\b~&=~\d_\a{}^\b\,\widetilde{\O}\\[5pt]
D_\a\widetilde{\O}\hskip4pt~&=~0~.
\end{split}\eeqnn

It is interesting to note also that
\beqnn
[D_\a, D_\bb]\;\O ~= -G_{\a\bb}\, \O,
\eeqnn
so the field strength, equivalently first Chern class, for the \K line bundle gives rise to the metric on the space of complex structures.

\subsection{Covariant derivatives for heterotic structures}\label{sec:derivhet}
This subsection motivates the  introduction of a connection on the moduli space of connections. Following 
Itoh~\cite{Itoh:1988}, for a fixed  $\ccX$, we introduce parameters $w^I$ describing the bundle deformations. Once we allow the gauge function to depend on the parameters, the transformation rules, for example for the derivatives of the gauge potential $\partial_I A$ and of the field strength $\partial_I F$, acquire extra terms that involve the derivatives of the gauge function. In order to restore the transformation properties to the expected form, we wish to define covariant derivatives with respect to the bundle parameters, $w^I$.  Later we will discuss how this derivative is generalised to all parameters. 

Under a $w$-dependent gauge transformation the derivative of the vector potential transforms according to the rule
\beqnn
\partial_I A~\to~\Ph(\partial_I A - \dd_A Y_I)\Ph^{-1}~~~\text{with}~~~
Y_I~=~\Ph^{-1}\partial_I\Ph~.
\eeqnn
We therefore introduce a new connection $\L_I$, which transforms in the form
\beq\label{LambdaRule}
\L_I~\to~\Ph \L_I \Ph^{-1} - \partial_I \Ph\, \Ph^{-1} ~=~\Ph (\L_I - Y_I)\Ph^{-1} ~,
\eeq
and a covariant derivative $D$, that is defined by
\beq
D_I A~=~\partial_I A - \dd_A \L_I~.
\label{CovA}\eeq
The covariant derivative now transforms homogeneously
\beqnn
D_I A~\to~ \Ph \,D_I A\,\Ph^{-1}~.
\eeqnn
Consider now the field strength. The partial derivative of $F$ is related to that of $A$ by
\beq
\partial_I F~=~ \dd_A \partial_I A~.
\label{PartialF}\eeq
The covariant derivative of the field strength should be defined by the relation
\beqnn
D_I F~=~\partial_I F + [\L_I, F]~,
\eeqnn
since, with this definition, the covariant derivative again transforms homogeneously
\beqnn
D_I F~\to~\Ph\, D_I F \,\Ph^{-1}~.
\eeqnn
From \eqref{PartialF}, we have
\beq\begin{split}
D_I F~&=~\dd_A(\partial_I A) +[\L_I, F] \\[3pt]
&=~\dd_A (\partial_I A) - \dd_A^2 \L_I \\[3pt]
&=~\dd_A(D_I A)~.
\end{split}\label{DF}\eeq
We may take the $(0,2)$ part of this last relation to find that
\beq
\delbA\big(D_I\cA\big)~=~0~.
\label{delbAexact}\eeq
It follows from \eqref{delbAexact} that $D_I\cA \in \Hend$. Just as for \eqref{CSCovDeriv}, the set of $D_i\cA$ constitute a choice of basis for $\Hend$. The basis means we can parameterise the deformation of the connection as
$$
\d\cA = \d w^i\, D_i \cA.
$$

The spin connection $\Th$ is a connection for the tangent bundle, which has structure group $\text{SU}(3)$, and the covariant derivative $D_\xi \Th$  is defined analogously to $D_\xi A$:
\beq
D_\xi \Th ~=~\partial_\xi \Th - \dd_\Th \Pi_\xi~,
\label{CovTheta}\eeq
where $\Pi_\xi$ is a connection analogous to $\L_\xi$. 

The moduli space is itself a complex manifold, and so we can introduce a complex structure giving the parameters holomorphic coordinates $w^I = (w^i, w^\jb)$.  
This also means we  can decompose $\L$ into its $(0,1)$ and $(1,0)$ parts
$$
\L~=~\l - \l^\dag~~;~~\l~=~\l_{\bar\jmath}\,\dd w^{\bar\jmath},
$$
and we can take
\beq
\l_{\bar\jmath}~=~-\partial_{\bar\jmath}\m\,\m^{-1}~.\label{eq:gaugePrePotential}
\eeq
As noted previously, the gauge prepotential $\m$ transforms according to the rule
\hbox{$\m\to \Ph \m \z$}, with $\z$ holomorphic. It follows that $\l_{\bar\jmath}$ transforms according to \eqref{LambdaRule}.

The connection $\L$ has a field strength
$$
\IF_{i\jb} = \del_{i} \L_{\jb}  - \del_{\jb} \L_{i} + [\L_i , \L_{\jb}].
$$
Similarly, we denote $\IG$ the field strength for $\Pi$: $\IG_{i\jb} = \del_{i} \Pi_{\jb}  - \del_{\jb} \Pi_{i} + [\Pi_i , \Pi_{\jb}]$.

The gauge potential $\A$ depends holomorphically on the bundle parameters in the following sense:
\beq
D_{\bar\jmath\,}\cA~=~0~.
\label{holomorphy}\eeq
This is manifest when  $\cA$ and $\l_{\bar\jmath}$ are written in terms of the gauge prepotential. Substituting \eqref{eq:gaugePrePotential} into $D_\jb \A$ we find
$$
D_{\bar\jmath\,}\cA~=~\partial_{\bar\jmath\,}\cA - \delb_{\!\cA}\l_{\bar\jmath}~=~
\partial_{\bar\jmath}\cA - \delb\l_{\bar\jmath} - [\cA,\l_{\bar\jmath}] ~=~ 0.
$$

\subsection{Variations of \K and complex structure}
We now allow  $\ccX$ to vary with parameters. These additional parameters derive from the complex structure and \K structure of $\ccX$. The construction of a derivative with respect to the \K parameters, $t^r$, is analogous to \eqref{CovA}:
$$
D_r A = \del_r A - \dd_A \L_r.
$$
Variation of the complex structure parameters presents special features, not present for the case of the bundle parameters or the \K-class parameters, owing to the fact that, even without depending in an explicit way on the complex structure, the gauge potential $\cA$ must inevitably vary owing to the fact that it is of type $(0,1)$ and so will mix with the $(1,0)$-part. Furthermore, the bundle needs to remain holomorphic and this further constrains the types of allowed deformations.

Before examining these considerations we pause  to review aspects of the theory of the variation of complex structure. 

Let us define $(0,1)$-forms with values in $\ccT_\ccX$, or equivalently tensors $\D_{\a\,\bar\n\bar\m}$ via the relation
$$
\pd{}{z^\a}\,\dd x^\m\,\Big|^{(0,1)}~=~\D_\a{}^\m~=~\D_{\a\,\bar\n}{}^\m\,\dd x^{\bar\n}~.
$$
Alternatively, the variation of complex structure can be described in terms of the variation of the holomorphic three form by noting that $\del_\a\O\in H^{(3,0)}\oplus H^{(2,1)}$ and writing
$$
\pd{\O}{z^\a}~=~k_\a\,\O + \ch_\a~;~~~
\ch_\a~=~\frac12 \ch_{\a\,\k\l\bar\n}\,\dd x^\k \dd x^\l  \dd x^{\bar\n}~.
$$
By performing the indicated differentiation we see that
\beq
\D_{\a\,\bar\n}{}^\m~=~\frac1{2\|\O\|^2}\, \chi_{\a\,\k\l\bar\n}\,\overline{\O}^{\,\k\l\m}~~~
\text{with}~~~
\|\O\|^2~=~\frac{1}{3!}\,\O_{\k\l\m}\overline{\O}^{\,\k\l\m}~.
\label{eq:DeltaChiEq}\eeq
The components of $\D_\a{}^\m$ are also directly related to the variations of the metric
\beq
\d g_{\bar\m\bar\n}~=~\D_{\a\,(\bar\m\bar\n)}\, \d z^\a~~~\text{and}~~~
\d g_{\m\n}~=~\D_{\bar\b\,(\m\n)}\, \d z^{\bar\b}~.
\label{eq:gDeltaEq}\eeq

We may write $\cA$ in terms of the total potential and the complex structure $J_m{}^n$
\beq
\cA~=~\dd x^m Q_m{}^n A_n~~~ \text{where} ~~~Q_m{}^n~=~\frac12\left( \d_m{}^n + \ii J_m{}^n\right)~.\label{eq:AQ}
\eeq
It follows that 
\beq
\d\cA~=~\frac{\ii}{2}\, \d J_m{}^n\, dx^m\, A_n + Q_m{}^n \d A_n dx^m ~.\label{eq:varA}
\eeq
The variations of $J$ are restricted by the the fact that $J^2=-1$ which has the consequence that the pure parts of the variation vanish, $\d J_\m{}^\n = 0$ and $\d J_{\bar\m}{}^{\bar\n} = 0$.
It is easy to see that the mixed terms of the variation are given by the important relations
\beq
\d J_{\bar\m}{}^\n 
~=~ 2\ii\, \d z^{\a}\,\D_{\a\,\bar\m}{}^\n~
~~~\text{and}~~~
\d J_\m{}^{\bar\n} ~=~-2\ii\, \d z^{\bar\b}\,\D_{\bar\b\,\m}{}^{\bar\n}.
\label{varJ}\eeq
The operator
$$
\delb~=~\dd x^m Q_m{}^n\, \partial_n
$$
undergoes a variation purely as a consequence of the implicit dependence on the complex structure: 
\beq
\left[\, \d,\,\delb \,\right] ~=~ -\d z^\a \D_\a{}^\m \partial_\m \,+\,
\d z^{\bar\a}\D_{\bar\a}{}^{\bar\m}\partial_{\bar\m}~, \qquad 
\left[\, \d,\,\del \,\right] ~=~ \d z^\a \D_\a{}^\m \partial_\m \,-\,
\d z^{\bar\a}\D_{\bar\a}{}^{\bar\m}\partial_{\bar\m}~.
\label{derivdelbar}\eeq

\subsection{Holotypical derivatives}\label{sec:holoD}
A partial derivative with respect to complex structure parameters of $\A$ consists of two parts:
$$
\del_\a \A ~=~  \D_{\a}{}^{\m} \cA^\dag_{\m} + \left(\partial_{\a} A\right)^{(0,1)}~.
$$
The first term arises from the fact $\A$ is a $(0,1)$-form, and so under a variation of complex structure a $(1,0)$-part is generated. One can think of it as coming from differentiating the projector $Q$ in \eqref{eq:AQ}. The second term $\left(\partial_{\a} A\right)^{(0,1)} $ contains any explicit dependence of $\A$ on complex structure, and derives from the fact the real form $A$ may depend on complex structure. The presence of these two terms suggests a refinement of the covariant derivative. 

The covariant derivatives of $A$ and $F$ with respect to complex structure, written in real form, are given by
$$
D_a A~=~\partial_a A -\dd_A\L_a~~~\text{and}~~~D_a F~=~\partial_a F + \big[\L_a, F\big]~.
$$

In order to write complex gauge covariant derivatives, similar to $D_i \cA$ described above for the bundle parameters, we introduce the {\em holotypical derivative}, denoted by~$\ccD$.  It  is defined via the relations
\beq\begin{split}
\ccD_\a\cA~&=~(D_\a A)^{(0,1)}~=~\partial_\a\cA - \D_\a{}^\m \cA^\dag_\m + \delbA \l^\dag_\a~, \\[5pt]
\ccD_{\bar\b}\cA~&=~(D_{\bar\b}A)^{(0,1)}~=~
\partial_{\bar\b}\cA - \D_{\bar\b}{}^{\bar\m} \cA_{\bar\m} - \delbA \l_{\bar\b}~=~0~,
\end{split}\label{CpxCovA}\eeq
where the vanishing of $\ccD_{\bar\b}\cA$ follows from \eqref{derivdelbar}. It follows from the definition 
that under a gauge transformation the holotypical derivative transforms in the desired form
\beq
\ccD_\a\cA ~\to~ \Ph\, \ccD_\a\cA\, \Ph^{-1}~.
\notag\eeq
Note that without the `extra term' $- \D_\a{}^\m \cA^\dag_\m$, this desirable property does not hold owing to the fact that, under a gauge transformation, a term $\delb\Ph$ appears and, as we have seen above, the $\delb$ fails to commute with $\del_\a$.

A further feature of the holotypical derivative is that it commutes with  decomposing forms into type: 
$\ccD_\a \A$ is manifestly a $(0,1)$-form. We will extend the holotypical derivative so as to act on $(p,q)$-forms below.

It will be useful later to define a holotypical derivative for all parameters  $y^\xi = (w^i, z^\a, t^\r)$.  We do this by extending $\D_\a$ to all parameters by setting $\D_i = \D_\r = 0$. This means 
$\ccD_i = D_i$ and $\ccD_\r = D_\r$. Note also that acting on a real form we have
\beq
\ccD_\x A ~=~\ccD_\x \cA -\ccD_\x \cA^\dag~=~ D_\x A~=~\ccD_\x\cA~.
\notag\eeq 
When the holotypical derivative acts on forms with tensor indices, it is defined to include a Levi-Civita symbol so that it transforms covariantly under diffeomorphisms. 
 
\subsubsection{The Atiyah constraint}
Considerations, analogous to those above, apply to the field strength $F$, which is of type $(1,1)$. Under a complex structure variation, $F$ would be expected to generate $(0,2)$ and $(2,0)$ parts. This observation was first made by Atiyah~\cite{Atiyah:1955} who noted that the continued vanishing of the $(0,2)$-part under a variation of complex structure is a condition on the allowed variations of the complex structure.

The condition that the field strength is of type $(1,1)$ is
$$
F_{mn}~=~2P_{[m}{}^p Q_{n]}{}^q\, F_{pq}~;\qquad 
2P_{[m}{}^p Q_{n]}{}^q ~=~  P_{m}{}^p Q_{n}{}^q - P_{n}{}^p Q_{m}{}^q
$$
On variation of $J\to J + \d J$,  the field strength can acquire a $(0,2)$-part owing to two processes: the implicit dependence, owing to the fact that $F$ is of type $(1,1)$, this gives rise to a term
\beq
-\D_\a{}^\m F_\m~,~~~\text{with}~~~F_\m~=~F_{\m\bar\n}\,\dd x^{\bar\n}
\notag\eeq 
that comes from varying the projectors using \eqref{varJ}. There is a second term that arises from the explicit dependence, since $F$ derives from $A$ and the (real) gauge potential can depend on the complex structure parameters. In virtue of \eqref{DF}, this yields a contribution
\beq
(D_\a F)^{(0,2)}~=~\delb_\cA(\ccD_\a \cA)~.
\notag\eeq
The sum of these two contributions must vanish if the vanishing of $F^{(0,2)}$ continues to hold. This yields the Atiyah constraint
\beq
\D_\a{}^\m F_\m~=~\delb_\cA(\ccD_\a \cA)
\label{AtiyahCond}\eeq
In other words, only variations 
such that the product $\D_\a{}^\m F_\m$ is trivial in cohomology maintain the condition that the bundle be holomorphic.

We can extend this equation for a general parameter variation,
\beq
\D_\xi{}^\m F_\m~=~\delbA (\ccD_{\xi}\cA)~.
\notag\eeq
since, for parameters that do not affect the complex structure, both sides of the equation vanish.
\subsubsection{Holotypical derivatives of $(p,q)$-forms}
Let $W = \sum_{p=0}^n W^{p,n-p}$ be an $n$-form.  Then, using the abbreviations
\begin{gather*}
P_{m_1\cdots m_n}^{i_1\cdots i_n}~=~P_{m_1}^{i_1}\cdots P_{m_n}^{i_n}~,\qquad
Q_{m_1\cdots m_n}^{i_1\cdots i_n}~=~Q_{m_1}^{i_1}\cdots Q_{m_n}^{i_n}~,\\[5pt]
\dd x^{m_1\cdots m_n}~=~\dd x^{m_1} \cdots  \dd x^{m_n}~,
\end{gather*}
we have
$$
W^{p,q}  =  \frac{1}{p!q!}\, P_{m_1\cdots m_p}^{i_1\cdots i_p} 
Q_{m_{p+1}\cdots m_{p+q}}^{j_1 \cdots j_q}  W_{i_1\ldots i_p j_1 \ldots j_q } \dd x^{m_1\cdots  m_{p+q}}~.
$$
Differentiating
\beq
\begin{split}
D_\a W^{p,q}~&= \\[5pt] 
&\hskip-30pt \frac{1}{p!q!}\left( p\, \D_{\a\, m_1}^{~~~~i_1} P_{m_2\cdots m_p}^{i_2\cdots i_p} 
Q_{m_{p+1}\cdots m_{p+q}}^{j_1\cdots j_q} 
 - q P_{m_1\cdots m_p}^{i_1\cdots i_p}\, \D_{\a\, m_{p+1}}^{~~~~j_1} 
 Q_{m_{p+2}\cdots m_{p+q}}^{j_2\cdots j_q} \right)
W_{i_1\ldots i_pj_1\ldots j_q} \dd x^{m_1\cdots  m_{p+q}} \\[5pt]
&~ +  \frac{1}{p!q!} \, P_{m_1\cdots m_p}^{i_1\cdots i_p} Q_{m_{p+1}\cdots m_{p+q}}^{j_1\cdots j_q}  
(D_\a W)_{i_1\ldots i_p j_1 \ldots j_q } \dd x^{m_1\cdots  m_{p+q}}\\[8pt]
&= \D_\a^{~\m}\, W_\m^{p-1,q} - \D_\a^{~\m}\, W_\m^{p,q-1} + (D_\a W)^{p,q}~,
\end{split}\label{eq:Wderiv}\raisetag{45pt}
\eeq
We use $D_\a$ to denote the covariant derivative to account for any gauge dependence of the real form $W$. 
In an analogous way we have
\beq
D_{\bar\b}W^{p,q}~=~\D_{\bar\b}{}^{\bar\n}\, W_{\bar\n}^{p-1,q} - 
\D_{\bar\b}{}^{\bar\n}\, W_{\bar\n}^{p,q-1} + (D_{\bar\b} W)^{p,q}~,
\label{eq:WderivAntiHol}\eeq
where we define
$$
W_k^{r,s} ~=~ \frac{1}{r!s!}\, W_{k, \m_1\cdots\m_r \nb_1\cdots\nb_s} 
\dd x^{\m_1\cdots\m_r\nb_1\cdots\nb_s},
$$
and understand $W_k^{r,s} = 0$ if $r$ or $s$ are negative or $r{+}s>n{-}1$.  The holotypical derivatives are then given by
\beq\begin{split}
\ccD_\a W^{p,q} ~&=~(D_\a W)^{p,q}~=~D_\a W^{p,q} - \D_\a^{~\m} W_\m^{p-1,q} + \D_\a^{~\m} W_\m^{p,q-1}\\[8pt]
\ccD_{\bar\b} W^{p,q} ~&=~(D_{\bar\b} W)^{p,q}~=~D_{\bar\b} W^{p,q} + 
\D_{\bar\b}{}^{\bar\n} W_{\bar\n}^{p-1,q} - \D_{\bar\b}{}^{\bar\n} W_{\bar\n}^{p,q-1}.
\end{split}\label{eq:CovariantDerivW}\eeq
The second expression is most easily derived by complex conjugation of the first. Note the change of signs that arises. The definitions above agree with those previously given for $\ccD_\a \A$. Furthermore for the holotypical derivative of $F^{(0,2)}$ we have
\beq
\ccD_\a F^{(0,2)}~=-\D_\a{}^\m F_\m~,
\notag\eeq
a result we have used in relation to the Atiyah constraint.

It is a straightforward check that 
\beq
\sum_{p=0}^n \ccD_\a W^{p,n-p} = D_\a W~,
\notag\eeq
so the holotypical derivative for a real $n$-form coincides with the covariant derivative.

Higher order holotypical derivatives follow by iteration of the first derivative. Thus the second order derivative 
$\ccD_\bb \ccD_\a W^{p,q}$, for example, is given by
\beq
\ccD_\bb \ccD_\a W^{p,q}~=~D_\bb (\ccD_\a W)^{p,q} + \D_\bb{}^{\bar\n} (\ccD_\a W)_{\bar\n}^{p-1,q}
- \D_\bb{}^{\bar\n} (\ccD_\a W)_{\bar\n}^{p,q-1}~.
\notag\eeq
The second and third terms, on the right, derive from the corresponding terms of the first derivative
\beq
(\ccD_\a W)_{\bar\n}^{p-1,q}~=~(\ccD_\a W^{p-1,q+1})_{\bar\n}^{p-1,q}~~~\text{and}~~~
(\ccD_\a W)_{\bar\n}^{p,q-1}~=~ (\ccD_\a W^{p,q})_{\bar\n}^{p,q-1}~.
\notag\eeq

We also have
\beq
\ccD_\bb \ccD_\a W^{p,q}~=~(D_\bb \ccD_\a W)^{p,q}~=~(D_\bb D_\a W)^{p,q}~.
\notag\eeq

The holotypical derivative should be extended to apply also to forms that have complex structure indices. Thus, acting on a form $W_\a^{(p,q)}$, say, the holotypical derivative should contain terms involving the holomorphic connection, which, on a \K manifold, coincides with the Christoffel connection, in the usual way.
\subsubsection{Holotypical derivatives of $H$ and $\o$}
Two particular forms that are of interest are $H$ and $\o$. From the supersymmetry relation $H= i (\del-\delb)\o$ and the fact that $\o$ is a $(1,1)$-form, we have that $H=H^{(2,1)}{+}H^{(1,2)}$. Using this, we list the various holotypical derivatives:
\begin{align}
 \ccD_\xi H^{3,0} ~&=~ 0\, , \notag\\[5pt]
 \ccD_\xi H^{2,1} ~&=~ \del_\xi H^{2,1} - \D_\xi{}^\m H_\m^{1,1}\, ,\notag\\[5pt]
 \ccD_\xi H^{1,2} ~&=~ \del_\xi H^{1,2} - \D_\xi{}^\m H_\m^{0,2} + \D_\xi{}^\m H_\m^{1,1}\, ,
 \notag\\[5pt]
 \ccD_\xi H^{0,3} ~&=~ \D_\xi{}^\m H_\m^{0,2}\, .
\label{eq:CovDerivH}
\intertext{The holotypical derivatives for the hermitian form $\o$ are }
\ccD_\xi \o^{2,0}~&=~0\, ,\notag\\[3pt]
\ccD_\xi \o^{1,1}~&=~ \del_\xi \o^{1,1} - \D_\xi{}^\m \o_\m^{0,1}\, ,\notag\\[3pt]
\ccD_\xi \o^{0,2}~&=~\D_\xi{}^\m \o_\m^{0,1}\, . \label{eq:OmegaExplicitComplexStruct}
\end{align}
with $\o^{0,1}_\m = \o_{\m\bar\n} \dd x^{\bar\n}$. 

Notice that, even though $\o$ is constrained to be of type $(1,1)$ and $H$ is thereby constrained so that 
$H=H^{2,1}{+}H^{1,2}$, it is still the case that
\beq
\ccD_\xi \o~=~\del_\xi \o~~~\text{and}~~~\ccD_\xi H~=~\del_\xi H~.
\notag\eeq
Note also that it is not asserted that $\ccD_\xi H^{0,3}$ and $\ccD_\xi \o^{0,2}$ vanish. We have
$\ccD_\xi H^{0,3}{=}(\del_\xi H)^{0,3}$, for example, and this quantity takes the value given in \eqref{eq:CovDerivH} so that $H^{0,3}$, evaluated with respect to the new complex structure, should vanish.

\subsubsection{Covariant and holotypical derivatives of $B$}\label{sec:derivB}
The $B$-field develops a gauge dependence at $\cO(\ap)$. 
Recall, the field strength $H$ is defined in \eqref{Hdef}, and $B$ transforms in way so that that $H$ is gauge invariant:
\beq
{}^\Phi B ~=~ B + \frac{\ap}{4}\Big( \tr \big(Y\! A - Z \Th\big) +  U\! - W \Big)~,
\label{eq:BTransf} 
\eeq
where $U$ and $W$ are such that $\dd U = \frac{1}{3} \tr (Y^3)$ and $\dd W = \frac{1}{3} \tr (Z^3)$. 

We wish to construct a covariant derivative for the $B$-field. This derivative will be chosen such that, under a gauge transformation, the transformation law for $D_M B$ is similar to that for $B$.
In this subsection, in order to avoid long expressions, we will compute derivatives including only the terms that relate to the gauge group. The parallel terms, that relate to Lorentz-frame rotations, will be added when stating final results.

First we define a quantity $U_M$ by setting\beq
U_M ~=~ \del_M U - \tr(Y^2 Y_M) +\dd\tr(A Y_M - \L_M Y)~.
\notag\eeq
Note that in virtue of the identity
\beq
\del_M Y ~=~ \dd Y_M + [Y, Y_M]
\notag\eeq
we have
\beq
 \dd U_M ~=~ \del_M\, \dd U - \frac{1}{3} \del_M \tr (Y^3)  ~=~ 0~.
\notag\eeq
We now take the covariant derivative of $B$ to be given by
\beq
D_M B = \del_M B -\frac{\ap}{4} \tr (\L_M\,\dd A) ~,
\label{eq:CovDerivB}
\eeq
With this choice, we have a gauge transformation law for~$D_M B$ that is parallel to the gauge transformation \eqref{eq:BTransf} for $B$: 
\beq
 {}^\Phi D_M B ~=~ D_M B + \frac{\ap}{4}\Big(\! \tr( Y D_M A ) + U_M \Big) ~.
\label{eq:covBTransf}\eeq

It is useful also to define a gauge invariant quantity $\ccB_M$ formed from $D_M B$
\beq
\ccB_M ~=~ D_M B + \frac{\ap}{4} \tr (A D_M A) -\dd b_M~,
\label{eq:curlyBdef}\eeq
with $\dd b_M$ an exact form. The exact form comes from the fact the physical quantity is $\dd B$, and so in writing $\ccB_M$ there is a corresponding  ambiguity. 
It is a simple exercise to note that $\del_M H$ is given by the expression
\beq 
\del_M H ~=~  \dd \ccB_M - \frac{\ap}{2} \tr ( D_M A\, F ) ~.
\label{eq:HDeriv}\eeq
and it follows immediately that ${}^\Phi (\del_M H) = \del_M H$, as it should.

As all forms above are real, we did not  need to discuss complex structure explicitly. However, when considering the $(p,q)$-component of the B-field, we need to introduce a holotypical derivative. This is constructed as before by projecting onto components:
\beq
 \ccD_\xi B^{p,q} ~=~ (D_\xi B)^{p,q} ~=~ ( \del_\xi B)^{p,q} - 
\frac{\ap}{4} \tr ( \dd A\, \L_\xi )^{p,q}~.
\label{eq:BCovDeriv0}\eeq

We will also have need for the mixed second order holotypical derivatives of the B-field. The form of the first derivative \eqref{eq:CovDerivB} suggests we take the following form for the second derivative
\beq
D_M D_N B ~=~  \del_M D_N B - \frac{\ap}{4} \tr ( \L_M\, \dd D_N A),
\label{eq:SecondDerivB}
\eeq
The gauge transformation property is  
\beq
 {}^\Phi D_M D_N B  ~=~   D_M D_N B +  
 \frac{\ap}{4}\Big(\! \tr ( Y  D_{M} D_N A)    + U_{MN} \Big)  ~,
\label{eq:BCovDeriv}\eeq
with
\beq
U_{MN}~=~\del_M U_N + \dd\tr\big( Y_M\, D_N A \big)~.
\notag\eeq

A short calculation, using this definition of $D_M D_N B$ yields the useful relation
\beq
\begin{split}
 [D_M, D_N] B &= -\frac{\ap}{4} \tr\Big( \dd A\, \IF_{MN} \Big)    + \frac{\ap}{4}\dd \tr \Big([\L_M,\L_N]A\Big). \end{split}\label{eq:DerivCommB}
\eeq 
As a consistency check, one can calculate
\beq
\big[ \del_M,\,\del_N \big] H~=~\dd \Big( \big[ D_M,\,D_N \big] B  + 
\frac{\ap}{4}\tr\big( \IF_{MN}\,\dd A \big) \Big)
\notag\eeq
and the right hand side vanishes, in virtue of the relation above.
\newpage
\subsection{Summary of derivatives and associated results}
We summarise these results in a table: 

\vskip15pt
\begin{table}[H]
\def\bigstr{\vrule height22pt depth12pt width0pt}
\def\medstr{\vrule height20pt depth11pt width0pt}
\def\smallstr{\vrule height17pt depth9pt width0pt}
\begin{center}
\begin{tabular}{|  l 
 |}
\hline
\medstr\hfil\large Some useful derivatives \\
\hline
\noalign{\vskip10pt}
\hline
\smallstr Real parameters $y^M = (w^I, z^a, t^r)$ \\ 
\hline\hline
\smallstr $D_M A~=~\partial_M A - \dd_A \L_M$ \\
\hline
\smallstr $\!\big[ D_M, D_N \big] A~=- \dd_A \IF_{MN}$ \\
\hline
\smallstr  $D_M F ~=~ \dd_A(D_M A)$ \\
\hline
\smallstr $D_M B ~=~ \del_M B -\frac{\ap}{4} \tr (\L_M\,\dd A - \Pi_M \dd \Th)$ \\
\hline
\smallstr $D_M D_N B ~=~  \del_M D_N B - 
\frac{\ap}{4} \tr\Big( \L_M\, \dd (D_N A) - \Pi_M \dd (D_N \Th) \Big)$ \\
\hline
\smallstr $\del_M H \,=\,  \dd \ccB_M - \frac{\ap}{2} \tr \big( D_M A\, F - D_M \Th\, R \big)\,;~~
\ccB_M =\, \ccD_\x B + \frac{\ap}{4} \tr \big( A D_\x A - \Th D_\x \Th \big) -\dd b_\x$ \\
\hline
\noalign{\vskip10pt}
\hline
\smallstr Holomorphic parameters $y^\xi = (w^i,z^\a,t^\r)$ \\ 
\hline\hline
\smallstr $\ccD_\xi \A ~=~ \partial_\xi\cA - \D_\xi{}^\m \cA^\dag_\m + \delbA \l^\dag_\xi$ \\
\hline
\smallstr $\ccD_{\eb}\cA~=~0$ \\ 
\hline
\smallstr $D_\eb D_\xi \cA ~=~ \delb_{\!\cA} \big(\IF_{\xi\eb}\big)$ \\
\hline
\smallstr $D_\xi F = \dd_A(D_\xi \A)$ \\ 
\hline
\smallstr $\delbA \big(\ccD_\x \cA\big) ~=~ \D_\xi{}^\m F_\m$ \\
\hline
\noalign{\vskip10pt}
\hline
\smallstr Gauge transformation of $B$ and its derivatives \\ 
\hline
\hline
\smallstr ${}^\Ph B~=~B + \frac{\ap}{4}\Big( \tr\big( Y A - Z \Th \big) + U - W \Big)$ \\
\hline
\smallstr ${}^\Ph D_M B~=~D_M B + \frac{\ap}{4}\Big(\tr\big(YD_M A - ZD_M \Th\big) + U_M - W_M\Big)$\\
\hline
\smallstr ${}^\Ph D_M D_N B~=~D_M D_N B + 
\frac{\ap}{4}\Big( \tr\big( Y D_M D_N A - Z D_M D_N \Th \big) + U_{MN} - W_{MN} \Big)$ \\
\hline
\end{tabular}
\capt{6.1in}{parameterDerivs}{A collection of results relating to the derivatives of important quantities with respect to parameters, together with the gauge transformation rules for $B$ and its first two derivatives.}
\end{center}
\end{table}
\newpage
\section{Supersymmetry relations between $\o$, $B$ and $A$}\label{sect:susyrelations}
Four-dimensional supersymmetry requires that the metric on the parameter space be \K.  In computing the metric from string theory it turns out the \K condition only follows if we make proper use of the supersymmetry identity 
\beq
\label{eq:SusyIdentity}
H ~=~ \dd^c \o~, \quad \dd^c \o ~=~ \half J_{m_1}{}^{n_1}J_{m_2}{}^{n_2} J_{m_3}{}^{n_3} (\del_{n_1}\o_{n_2n_3})\, \dd x^{m_1} \dd x^{m_2} \dd x^{m_3}~.
\eeq
This implies relations between first and second order variations of the hermitian form $\o$, the B-field and the gauge field $A$.  These relations are essential to constructing a \K moduli space metric and we aim to calculate these relations in this section. We start by reconciling the relations \eqref{eq:CovDerivH} and 
\eqref{eq:OmegaExplicitComplexStruct}, for the holotypical derivatives of $H$ and $\o$, with the supersymmetry relation \eqref{eq:SusyIdentity} above.

\subsection{First order  supersymmetry relations} 
The manifold $\ccX$ is complex, and so the Nijenhuis tensor vanishes
\beq
J_{[m_1}{}^{n_1}  \del_{|n_1|} J_{m_2]}{}^{n_2} -   J_{n_1}{}^{n_2}  \del_{[m_1} J_{m_2]}{}^{n_1} ~=~ 0~,\notag
\eeq
with the consequence that relation \eqref{eq:SusyIdentity} can be simplified
\beq
H ~=~ J^m \del_m \o - (\dd J^m)\, \o_m~;~~~J^m~=~J_n{}^m \dd x^n~.\label{eq:dcoSusy}
\eeq
For fixed complex structure the second term vanishes leaving what is commonly written as the supersymmetry condition
$$
H = \ii (\del - \delb )\,\o~.
$$
While the second term vanishes for fixed complex structure, it is non-vanishing when the complex structure of 
$\ccX$ is varied. So, it needs to be kept as it plays a role in the discussion of moduli.
 
Taking this into account and using \eqref{varJ} in differentiating \eqref{eq:dcoSusy} we have
$$
\del_\x H ~=~ \ii (\del-\delb) \del_\x \o + 2\ii \D_\x{}^\m (\del_\m \o - \del \o_\m)   - 2\ii \del( \ccD_\x \o^{0,2}) ~.
$$
Projecting onto type, we find 
\beq
\begin{split}
\ccD_\xi H^{3,0} ~&=~\, 0~,\\[3pt]
\ccD_\xi H^{2,1} ~&=~\, \ii\, \del \ccD_\xi \o^{1,1}~,\\[3pt]
\ccD_\xi H^{1,2} ~&=  - \ii \del \ccD_\x \o^{0,2} - \ii \delb \ccD_\x \o^{1,1} + 2\ii \D_\x{}^\m (\del_\m \o - \del \o_\m)  ~,\\[3pt]
\ccD_\xi H^{0,3} ~&=- \ii\, \delb \ccD_\xi \o^{0,2}~,
\end{split}\label{eq:HCovDeriv2}
\eeq
which we may compare with \eqref{eq:CovDerivH}. In writing these relations, we have used the fact that $\ccD_\x \o^{2,0}=0$, which we have from
\eqref{eq:OmegaExplicitComplexStruct}. We may also eliminate reference to $\ccD_\x \o^{0,2}$ from these relations since we have also that 
$\ccD_\x \o^{0,2} = \D_\x{}^\m \o_\m{}^{0,1}$. Although we will need only the first two relations above, note the consistency between the last relation and the last relation of \eqref{eq:CovDerivH}:
\beq
\ccD_\xi H^{0,3} ~= -\ii\, \delb(\D_\x{}^\m \o_\m^{0,1})
~=~\ii\, \D_\x{}^\m\, \delb\o_\m^{0,1}~=~\D_\x{}^\m H_\m{}^{0,2}~,
\notag\eeq
where the second equality uses the fact that $\delb\D_\x{}^\m = 0$.

We have been exploring the relations between the variations of $H$ and $\o$. The variations of $H$ are related also to those of $B$ and $A$, as for example, in \eqref{eq:HDeriv}.
On decomposing \eqref{eq:HDeriv} into type we have
\beq
\ccD_\x H^{p,q}~=~\del\ccB_\x^{p-1,q} + \delb\ccB_\x^{p,q-1} - 
\frac{\ap}{2} \tr\!\big(\ccD_\x\cA\, F^{p,q-1}\big)~.
\label{eq:varHByType}\eeq
We compare this with \eqref{eq:HCovDeriv2}, starting with the $(3,0)$ and $(0,3)$ parts.
The $(3,0)$ part yields
\beqnn
\del \ccB_\x^{2,0}  ~=~ 0~.
\eeqnn
As we are dealing with CY manifolds, $h^{2,0} = 0$ and so
\beq
\label{eq:ExactPeiceBfield}
\ccB_\x^{2,0} ~=~ \del \b_\xi^{1,0}, 
\eeq
with $\b_\xi^{1,0}$ a $(1,0)$-form. 

The $(0,3)$ part of \eqref{eq:varHByType} yields
\beq
\delb\big( \ccB_\x^{0,2} + \ii \ccD_\x \o^{0,2} \big)~=~0~.
\notag\eeq

Using   \eqref{eq:ExactPeiceBfield}, the $(2,1)$-component of \eqref{eq:HCovDeriv2}  is
\beq
\del \Big( \ccB_\x^{1,1} - \ii \ccD_\xi \o^{1,1} - \delb \b_\xi^{1,0} \Big) ~=~ 0~.
\label{eq:Closedform1}
\eeq
It follows that we can write
\beq
\ccB_\x^{1,1} - \ii \ccD_\xi \o^{1,1}~=~\g_\x^{1,1} + \del\a_\x^{0,1} + \delb \b_\xi^{1,0}~,
\label{eq:OmBRelation}\eeq
where $\g_\xi^{1,1}$ is $\dd$-closed $(1,1)$-form. 
This is a straightforward consequence of the $\del\delb$-lemma, and we pause briefly to explain why. 

For a $\del$-closed form $\ph$, set $\s=d\ph=\delb\ph$. Then $\s$ is 
$\dd$-exact. The $\del\delb$-lemma states that there is a form $\a$ such that $\s=\del\delb\a$. Now, set 
$\ph =  \ccB_\x^{1,1} - \ii \ccD_\xi \o^{1,1} - \delb \b_\xi^{1,0}$   and
$\g_\x=\ph_\x - \del\a_\x$. We see that $\del\g_\x=\delb\g_\x=0$, so $\g_\x$ is $\dd$-closed, as promised.

Recall that $\ccB_\x$ is defined by \eqref{eq:curlyBdef} up to an exact two-form 
$\dd b_M$. If we wish, we can use this freedom to remove the exact components from \eqref{eq:ExactPeiceBfield} and \eqref{eq:OmBRelation}, and also remove an exact part from $\g_\x^{1,1}$, and so leave us with the relations
\beq
\ccB_\x^{2,0}~=~0 ~~~\text{and}~~~\ccB_\x^{1,1} - \ii \ccD_\xi \o^{1,1}~=~\g^{1,1}_\x~,
\notag\eeq
with $\g^{1,1}_\x$ harmonic, say.
We will see later that we can also remove $\g^{1,1}_\x$ via a suitable change of coordinates on the parameter space so that
\beq
\label{eq:NewSusyRelationB}
\ccB_\x^{1,1} - \ii \ccD_\x \o^{1,1} ~=~ 0~.
\eeq 
This is the generalisation of the special geometry relation $\del_\r\, (B^{1,1} - \ii \o^{1,1}) = 0$ to include the first order $\ap$-corrections. 

Note however that we cannot also remove the exact piece from the $(0,2)$ part so we are left with
\beq
\ccB_\x^{0,2} + \ii \ccD_\x \o^{0,2}~=~\delb \k_\x^{0,1}~.
\label{eq:zerotwopart}\eeq
with $\k_\x^{0,1}$ a $(0,1)$ form.

The $(1,2)$ component of \eqref{eq:varHByType} yields the relation
\beq
\delb (\ccB_\x^{1,1} + \ii \ccD_\x\o^{1,1}) + \del (\ccB_\x^{0,2} + \ii \ccD_\x \o^{0,2}) ~=~ 2\ii \D_\x{}^\m (\del_\m \o - \del \o_\m) + \frac{\ap}{2} \tr( \ccD_\a \A F)~. 
\label{eq:21relation}\eeq

We summarise the first order relations as: 
\beq
\begin{split}
\ccB_\x^{2,0}  ~=~ \del \b^{1,0},~~&~~~~~ \ccD_\x\o^{2,0} ~=~ 0~,\\[3pt]
\ccB_\x^{0,2} + \ii \ccD_\x \o^{0,2} ~&=~ \delb \k_\x^{0,1}~, \\[3pt] \ccB_{\x}^{1,1} - \ii \ccD_\x \o^{1,1} &=~ \Big(\g_\x + \dd\left( \a^{0,1}_\x + \b^{1,0}_\x\right)\Big)^{1,1}~,\\[3pt]
\delb (\ccB_\x^{1,1} + \ii \ccD_\x\o^{1,1}) + \del (\ccB_\x^{0,2} + \ii \ccD_\x \o^{0,2}) ~&=~ 2\ii \D_\x{}^\m (\del_\m \o - \del \o_\m) + \frac{\ap}{2} \tr( \ccD_\a \A F)~, 
\end{split}\label{eq:cBrelations1}
\eeq
where $\g_\xi^{1,1}$ is $\dd$-closed $(1,1)$-form. As discussed in sections to come, in $\ap$-perturbation theory, $\ccB_\x^{0,2} = \ccD_\x \o^{0,2} = \cO(\ap)$ when appropriately gauge fixed. 

\subsection{Second order relations}

Let us turn now to second order holotypical derivatives of $\o$. They are
\beq
\begin{split}
 \ccD_\eb \ccD_\xi \o^{2,0} ~&=~   \D_\eb^{~~\nb} (\del_\xi \o)^{1,0}_{\nb}~,\cr
 \ccD_\eb \ccD_\xi \o^{1,1} ~&=~ \del_\eb (\ccD_\xi \o^{1,1})  + 
 \D_\eb^{~~\nb} (\del_\xi \o)^{0,1}_{\nb} - \D_\eb^{~~\nb} (\del_\xi \o)^{1,0}_{\nb}~,\cr
 \ccD_\eb \ccD_\xi \o^{0,2} ~&=~\del_\eb (\ccD_\x\o^{0,2})  - \D_\eb^{~\nb} (\del_\x\o)^{0,1}_\nb~.
\end{split}
\notag\eeq
A useful consistency check is that 
\beq
\ccD_\eb \ccD_\xi \o^{1,1} + \ccD_\eb\ccD_\xi \o^{2,0} + \ccD_\eb\ccD_\xi \o^{0,2} ~=~
\del_\eb (\ccD_\xi \o) ~=~ \del_\eb \del_\xi \o~.
\notag\eeq  

There is also an identity, that we will not need in the following but which we note in passing, that arises from the observation 
$[\ccD_\x, \ccD_\eb] \o^{p,q} = \big( [\del_\eb,\del_\x]\o \big)^{p,q} = 0$. 
The three choices of $(p,q)$ lead to the single identity:
\beq
\ccD_\x\big( \D_{\bar\eta}{}^{\bar\n} \big)\, g_{\s\bar\n} \dd x^\s ~=~0~.
\notag\eeq

Returning to our discussion, we now assume the coordinates have been appropriately chosen so that \eqref{eq:NewSusyRelationB} holds. Differentiating this identity gives a second order relation:
\beq
\begin{split}
0~=~ \del_\eb \Big(\ii \ccD_\x \o^{1,1} - \ccB^{1,1}_\x  \Big) ~&=~ \ii \ccD_\eb \ccD_\x \o^{1,1} - \ccD_\eb \ccB_\x^{1,1} - \D_\eb^{~~\nb} \Big(\ii  \ccD_\x \o_{\nb\rb}    - \ccB_{\x\,\nb\rb}  \Big)\dd x^\rb~. \\
\end{split}\nonumber
\eeq
Define
\beq\label{eq:defUp}
\U^{0,2}_\x = \ccB^{0,2}_\x - \ii \ccD_\x \o^{0,2}~.
\eeq
Notice that $\U^{0,2}_\x$ and $\g^{1,1}_\x$ are distinct quantities. 
We find
\beq
\ii \ccD_\eb \ccD_\x \o^{1,1} = \ccD_\eb \ccB_\x^{1,1} - \D_\eb^{~~\nb}\, \U_{\x\,\nb\rb}\,\dd x^\rb~.
\notag\eeq
It is important to keep in mind that
$
\U_\x^{0,2}  = \cO(\ap).
 $
 We will have need for the quantity $ \ii \{\ccD_\eb, \ccD_\x \}\o^{1,1}$, and it is related to the curl of $\ccB_\x$:
\beq
\begin{split}
\ii \{ \ccD_\eb, \ccD_\xi \}\,\o^{1,1} ~&=~ 2 \ccD_{[\eb} \ccB_{\x]}^{1,1}  
-\D_{\eb}^{~~\nb} \U^{0,1}_{\x\,\nb} + \D_\x^{~~\m}\U_{\eb\,\m}^{1,0} .
\end{split}\label{eq:anticommOm}
\eeq

The curl of $\ccB_\x$ derives from
\begin{align}
\notag \del_{[M} \ccB_{N]} ~&=~ \frac{\ap}{4} \tr \big(D_{[M} A D_{N]} A - \IF_{MN} F   \big) \, + \\[3pt]
&\hskip50pt\dd \left[ -\del_{[M} B_{N]} + \frac{\ap}{4}  \tr \Big( \L_{[M} D_{N]} A + 
A \del_{[M} \L_{N]}  \Big)\right]~. \label{eq:curlB}\\ 
\intertext{Hence,}
\notag \ii \{ \ccD_\x, \ccD_\eb \}\,\o^{1,1}~&=~
\frac{\ap}{2} \tr \big(D_\x \A D_{\eb} \A^\dag + \IF_{\x\eb} F^{1,1}\big) 
-\D_{\eb}^{~~\nb} \U^{0,1}_{\x\,\nb} + \D_\x^{~~\m}\U_{\eb\,\m}^{1,0}\,  + \\[3pt]
&\hskip50pt \left( \dd \Big[2\del_{[\x} B_{\eb]} - \frac{\ap}{2}  \tr \Big( \L_{[\x} D_{\eb]} A + 
A \del_{[\x} \L_{\eb]}  \Big)\Big] \right)^{1,1}.\label{eq:anticommOm2}
\end{align}
The $\dd$-exact piece does not play a role in the calculations to come. 
\newpage
\section{The parameter space metric}\label{sect:KahlerMetric}
\vskip-5pt
\subsection{The \K potential}
We  come to computing the parameter space metric. We compute it in two ways: the first is by computing the metric deriving from a \K potential which we propose with some prescience. The second is to dimensionally reduce $\ap$-corrected heterotic supergravity. The two methods agree as we show in the next section. 

We propose a \K that describes the $\ap$-corrected moduli space metric. It is remarkably similar to the special geometry \K potential, in which the \K form is replaced by the $\ap$-corrected hermitian form:
\beq
 K = K_1 + K_2 =  -\log\left( \ii \!\int\! \O\, \Ob \right) - \log\left( \frac{4}{3} \int \o^3\right)~. \label{eq:KahlerPotential}
\eeq
Although it is remarkably similar to the special geometry \K potential, in the derivation of the moduli space metric and \K potential, no assumptions are made about special geometry. The fact we arrived at such a similar \K potential is a surprising conclusion from our calculation. 
 
In this section we compute the metric using the results constructed in the previous two sections. The answer agrees with known mathematics literature in the situation with the CY is fixed. It also agrees with the answer we get from dimensionally reducing $\ap$-corrected supergravity in the next section. We conclude this is the \K parameter space metric and \K potential as dictated to us by $\ap$-corrected supergravity.

The first term, $K_1$, gives the complex structure  metric:
\beq
G^0_{\a\bb}~=~ \del_\a\del_\bb K_1 ~=~  \frac{1}{4V} \int \del_\a g_{\mb\nb} \del_\bb g^{\mb\nb} \star 1~= - \frac{\ii}{V\norm{\O}^2} \int \chi_\a \star \chi_\bb.\label{eq:SGCS}
\eeq
The second term $K_2$ contains all the $\ap$-corrections. Differentiating twice
\beq
\del_\xi \del_\eb K_2 ~=~ \frac{1}{V} \int \del_\xi \o \star \del_\eb \o- \frac{1}{2V} \int \o^2 \del_\xi \del_\eb \o~.\label{eq:KahlerPotMetric1}
\eeq

We need  to turn these terms into appropriate holotypical derivatives in order to  express the metric in gauge invariant quantities that reflect the physical moduli fields that arise in the dimensional reduction. The first term uses
$$
\del_\x \o ~=~ \ccD_\xi \o^{1,1} + \ccD_\xi \o^{0,2}~.  
$$
 For the second, we use $\o$ is a $(1,1)$-form and so 
 $$
 \o^2 \del_\x\del_\eb \o ~=~ \o^2 \ccD_\x\ccD_\eb \o^{1,1} ~=~ \frac{1}{2} \o^2 \{ \ccD_\x, \ccD_\eb\} \o^{1,1}.
 $$ 
 The second equality follows from $[\del_\x,\del_\eb] \o = 0$.

Returning to the \K potential,
\beqnn
\begin{split}
\del_\xi \del_\eb K_2~ 
&=~ \frac{1}{V} \int( \ccD_\xi \o^{1,1} + \ccD_\x \o^{0,2} ) \star (\ccD_\eb \o^{1,1} + \ccD_\eb \o^{2,0} ) - \frac{1}{4V} \int \o^2  \{\ccD_{\x}, \ccD_{\eb}\} \o^{1,1} \\[10pt]
&=~ \frac{1}{V}\!\! \int\!\!\left( \ccD_\xi \o^{1,1} \star \ccD_\eb \o^{1,1}  + \ccD_\x \o^{0,2} \star \ccD_\eb \o^{2,0} \right) + \frac{\ii}{4V} \!\!\int\!\!  \o^2 \ii \{\ccD_{\x}, \ccD_{\eb}\} \o^{1,1}~.
 \end{split}
\eeqnn
In the second term $\ccD_\x \o^{0,2} \star \ccD_\eb \o^{2,0}$ is $\cO(\ap^2)$. 
For the third term we use  \eqref{eq:anticommOm2}:
\beq
 \frac{\ii}{4V} \int \o^2 \ii \{\ccD_{\x}, \ccD_{\eb}\} \o = 
 \frac{\ii}{4V} \int \o^2\Big[ \frac{\ap}{2} \tr \ccD_\x \A \ccD_\eb \A^\dag  + \D_\x^{~~\m} \U_{\eb\,\m}^{1,0} - \D_\eb^{~~\nb}\U_{\x\,\nb}^{0,1}\Big]~,
\notag\eeq
where $\U_\x$ is defined in \eqref{eq:defUp}, and also
\beq
\half \o^2 \D_\eb^{~~\nb} \U_{\x\,\nb}^{0,1} 
~=~  \ccD_\eb \o^{2,0} \star \ccB_\x ^{0,2}  -\ii \ccD_\eb \o^{2,0} \star \ccD_\x\o^{0,2} 
~=~ \cO(\ap^2)~. 
\notag\eeq
In this way we see that
\beq
\frac{\ii}{4V} \int \o^2 \ii \{\ccD_{\x}, \ccD_{\eb}\} \o  ~=~ 
\frac{\ii \ap}{8V} \int \o^2 \tr \ccD_\x \A \ccD_\eb \A^\dag~.   
\notag\eeq
Hence, 
\beq
\del_\xi \del_\eb K_2 ~=~ 
\frac{1}{V}\!\! \int\!\! \ccD_\xi \o^{1,1} \star \ccD_\eb \o^{1,1}  +  \frac{\ii \ap}{8V} \int \o^2 \tr \ccD_\x \A \ccD_\eb \A^\dag 
\notag\eeq

Including the complex structure special geometry metric \eqref{eq:SGCS} we get
\beq
\dd s^2 = 2G^K_{\xi\eb}\, \dd y^\xi \dd y^\eb + 2G^0_{\a\bb}\, \dd z^\a \dd z^\bb~,
\label{eq:MetricModuli1}
\eeq
where $\dd y^\xi = \{ \dd z^\a, \dd t^\r, \dd w^i\}$. So when we choose complex coordinates on $\cM$ so that \eqref{eq:NewSusyRelationB} holds, a choice naturally handed to us by string theory as shown in the next section, we find the \K potential exactly gives the metric  \eqref{eq:FinalMetric1}-\eqref{eq:FinalMetric2} arising from the dimensional reduction.

The upshot is that the complex structure metric $G^0_{\a\bb}$ is unchanged in $\cO(\ap)$, while the  complexified \K metric $G^K_{\x\eb}$ is corrected, and, as written above, implicitly includes $\ap$-corrections. The complex structure metric can still, as is the case of special geometry, be written as a metric on the cohomology classes.  This is not obviously the case for the metric $G^K_{\x\eb}$, we intend to return to this point in future work. 

\subsection{Cohomological description of the parameter space}
\vskip-5pt
The structure of the parameter space at order $\ap$ is difficult to describe owing to the mixing between the complex, hermitian and bundle structures.  However, in \cite{delaOssa:2014cia, Anderson:2014xha, delaOssa:2015maa}, it was shown that the moduli of the Strominger/Hull system for four dimensional compactifications with $N=1$ supersymmetry, can be recast as a holomorphic structure $\overline D$ on an extension bundle $\cQ$ on $\ccX$, and that the moduli of these compactifications are given by the deformations of the holomorphic structure, that is, by elements in $H^1_{\raisebox{-2pt}{$\scriptstyle\overline D$}}(\cQ)$.  In this subsection we summarize the results of these papers following mainly the point of view of \cite{delaOssa:2014cia}.

For fixed $\ccX$,  the only parameters $w^i, w^\jb$ are those deforming the gauge connection such that Hermitian-Yang-Mills equation is preserved, as described by Kobayashi and Itoh \cite{Kobayashi:1987, Itoh:1988}.  If we allow the bundle and $\ccX$ to vary simultaneously, it is natural to look at the  deformation theory of the total space of the bundle. For fixed hermitian structure, this was described by Atiyah \cite{Atiyah:1955}, and in the context of heterotic string theory on a \cym by \cite{Anderson:2011ty}. Defining an extension bundle $Q_1$  via the short exact sequence:
\beq
\xymatrix{
0 \ar@{>}[r] &{\rm End}(\ccE)  \ar@{>}[r]  & Q_1 \ar@{>}[r]^{\pi_1}& \ccT_\ccX   \ar@{>}[r] & 0}~,
\notag\eeq
where $\pi_1: \ccE \to \ccX$ is the canonical projection, one can show that there is a holomorphic structure on this bundle which precisely describes the equations for moduli.  This induces a long exact sequence of cohomology, and complex structure deformations of the total space correspond to the cohomology group  $H^1(\ccX, Q_1)$. When $\ccE$ is a deformation of $\ccT_\ccX$ one finds
\beq
H^1(\ccX,Q_1) = H^1(\ccX,\ccT_\ccX) \oplus \Hend~.\label{CSDefStandardEmbedding}
\eeq
 In general though, the complex structure moduli of $\ccX$ are reduced
 \beq
 H^1(\ccX,Q_1) =  \Hend\oplus \ker\!{\cF}~,
 \label{AtiyahCohomology}
 \eeq
 where $\cF: H^1(\ccX, \ccT_\ccX)\to H^2(\ccX, {\rm End}\, \ccE)$ is the Atiyah map on cohomologies defined by
 \beq
 \cF(\Delta) = F_{\mu\bar\nu}\dd x^\nu\wedge\Delta^\mu~.
 \notag\eeq
 Note that the condition that $\Delta\in\ker\!\cF\subseteq H^1(\ccX,\ccT_\ccX)$ is the Atiyah constraint  \eqref{AtiyahCond}.

This calculation needs to be generalised to include the deformations of the hermitian structure, accounting simultaneously for the Yang-Mills equation and the anomaly cancelation condition.   In order to do this, we define an extension the bundle $\cQ$ by the short exact sequences
\beq
\xymatrix{
0 \ar@{>}[r] &\ccT^{\raisebox{1pt}{$*$}}_\ccX  \ar@{>}[r]  & \cQ \ar@{>}[r]^{\pi}& Q_2   \ar@{>}[r] & 0}~,
\notag\eeq
where
\beq
\xymatrix{
0 \ar@{>}[r] &{\rm End}(\ccT_\ccX)  \ar@{>}[r]  & Q_2 \ar@{>}[r]^{\pi_2}& Q_1   \ar@{>}[r] & 0}~.
\notag\eeq
The anomaly cancellation condition induces a holomorphic structure $\overline D$ on $\cQ$ and the moduli of the Strominger/Hull system is then given by the elements of
the cohomology $H^1_{\raisebox{-2pt}{$\scriptstyle\overline D$}}(\cQ)$.  The extension $Q_2$ of $Q_1$ by ${\rm End}\,\ccT_\ccX$ is necessary to enforce 
the connection on the tangent bundle appearing in the anomaly cancelation condition to be an  instanton \cite{Hull:1986kz, Ivanov:2009rh, Martelli:2010jx}. In fact, this is  needed to satisfy the equations of motion.  We do not give here the derivation of the holomorphic structure $\overline D$ on $\cQ$ nor the derivation of the cohomology groups corresponding to the moduli space.  The result however is that the moduli for the heterotic structure correspond to elements of the cohomology group
\beq
H^1_{\raisebox{-2pt}{$\scriptstyle\overline D$}}(\cQ) = H^1(\ccX,\ccT^{\raisebox{1pt}{$*$}}_\ccX) \oplus \ker\!\cH~,\qquad \ker\!\cH\subseteq H^1(\ccX, Q_2)~,
\label{AllModuli}\eeq
where
\beq
H^1(\ccX, Q_2) = H^1(\ccX, {\rm End}\,\ccT_\ccX)\oplus \Hend\oplus (\ker\!\cF\cap\ker\!\cR)~.
\label{HolModuli}\eeq
The first factor in \eqref{AllModuli} corresponds to complexified $\ap$-corrected hermitian moduli. The second factor contains a map $\cH: H^1(\ccX, Q_2)\to H^2(\ccX, \ccT^{\raisebox{1pt}{$*$}}_\ccX)$ defined by
\beq
\cH_\mu(\alpha,\kappa,\Delta) = H^{(2,1)}_{\mu\nu\bar\rho}\,\dd x^{\bar\rho}\wedge\Delta^\nu 
- \frac{\ap}{4\phantom{`}}\,\Big(\tr (F_{\mu\bar\nu}\dd x^{\bar\nu}\wedge\alpha)
- \tr (R_{\mu\bar\nu}\dd x^{\bar\nu}\wedge\kappa)
\Big)~,
\notag\eeq
where $\alpha$ is a $(0,1)$-form with values in ${\rm End}\, \ccE$ and $\kappa$ is a $(0,1)$-form with values in ${\rm End}\, \ccT_\ccX$.
There is a subtlety in that the parameters in \eqref{HolModuli} corresponding to  $H^1(\ccX, {\rm End}\, \ccT_\ccX)$ are not physical and can be removed by field redefinitions
\cite{delaOssa:2014msa}. The map $\cR$ in \eqref{HolModuli} is the Atiyah map appropriate for the deformations of the holomorphic tangent bundle.
Finally, we remark that the same results for the moduli problem of heterotic structures was obtained in \cite{delaOssa:2015maa} from first and second order deformations of a heterotic superpotential. 

The relation between the discussion in this sub-section to the parameters in this paper is as follows:  
\begin{itemize}
\item 
$z^\a$ denote parameters corresponding to deformations of the complex structure of $\ccX$ in $\ker\!\cF$ which are also in $\ker\!\cH$; 
\item 
$w^i$ denote parameters corresponding to those elements in $\Hend$ which are in~$\ker\!\cH$;
\item 
$t^\r$  are hermitian parameters corresponding to $H^1(\ccX,\ccT^{\raisebox{1pt}{$*$}}_\ccX)$.
\end{itemize}

\newpage
\section{Dimensional reduction of $\ap$-corrected heterotic supergravity}
\vskip-5pt
\subsection{Preliminaries}
In this section we dimensionally reduce heterotic supergravity on a Calabi--Yau manifold to determine the K\"ahler metric for bundle moduli. The heterotic action is fixed by supersymmetry up to $\cO(\alpha'^2)$. The action in string frame takes a particularly nice form when an appropriate choice of connection is made \cite{Bergshoeff:1989de,Bergshoeff:1988nn}:
\begin{equation}
S = \frac{1}{2\kappa_{10}^2} \int\! \dd^{10\,}\! X \sqrt{g_{10}}\, e^{-2\Phi} \Big\{ \cR -
\half |H|^2  + 4(\del \Phi)^2 - \frac{\alpha'}{4}\big( \tr |F|^2 {-} \tr |R(\Theta^+)|^2 \big) \Big\} + \cO(\alpha'^3),
\label{eq:10daction}
\end{equation}
Our notation is such that  $\m,\n,\ldots$ are holomorphic indices along $\ccX$ with coordinates $x$; $m,n,\ldots$ are real indices along $\ccX$; while $e,f,\ldots$ are spacetime indices corresponding to spacetime coordinates $X$.   The 10D Newton constant is denoted by $\kappa_{10}$,
\hbox{$g_{10}=-\det(g_{MN})$}, $\Phi$ is the 10D dilaton, $\cR$ is the Ricci scalar evaluted using the Levi-Civita connection and $F$ is the Yang--Mills field strength with the trace taken in the adjoint of the gauge group. 

We define a pointwise inner product on $p$-forms by
$$
\langle S,\, T\rangle~=~ 
\frac{1}{p!} \, g^{M_1 N_1} \ldots g^{M_p N_p}\, S_{M_1\ldots M_p} \,T_{N_1 \ldots N_p}
$$
and take the $p$-form norm as
$$
|T|^2 ~=~\langle T,\, T\rangle~.
$$
Thus the curvature squared terms correspond to
$$
\tr |F|^2 = \half \tr F_{MN} F^{MN}~~~\text{and}~~~ \tr |R(\Th^+)|^2 ~=~
\half \tr R_{MNPQ}(\Theta^+) R^{MNPQ}(\Theta^+)~,
$$
where the Riemann curvature is evaluated using a twisted connection
$$
\Theta^\pm_M = \Theta_M \pm \half H_M,
$$
with $\Theta_M$ is the Levi-Civita connection.
The definition of the $H$ field strength and its gauge transformations are given in \sref{sec:BHtransfs}.

The equations of motion, correct to second order in $\ap$, are given by
\begin{equation}\begin{split}
&  \cR - 4(\nabla \Phi)^2 + 4 \nabla^2 \Phi - \half |H|^2 - \frac{\alpha'}{4} \big(\tr |F|^2 - \tr |R|^2\big) ~=~0~,\\[3pt]
& \cR_{MN}+ 2 \nabla_M \nabla_N\Phi - \frac{1}{4} H_{MAB} H_N{}^{AB} -
\frac{\alpha'}{4} \Big( \tr F_{MP} F_N{}^P - R_{MPAB}(\Th^+)R_N{}^{PAB}(\Th^+)\Big)  ~=~ 0~,\\[6pt]
&\nabla^M(\ee^{-2\Phi} H_{MNP}) ~=~0~,\\[8pt]
&\ccD^{-\,M} (\ee^{-2\Phi} F_{MN}) ~=~0~,
\end{split}\raisetag{45pt}\label{EOM}\end{equation}
\vskip5pt
here $\ccD^- = \nabla^- + [A,\cdot]$,  with $\nabla^-$ computed with respect to the $\Th^-$ connnection, and $\cR_{MN}$ is the Ricci tensor.

\subsection{Small deformations of the Calabi--Yau background}
We start with a parameter space $\cM$ with real coordinates $y^M$ for a family of heterotic structures 
$(\ccX, \ccE, H)$.  Supersymmetry tells us that $\cM$ is a complex manifold and so there exists a complex structure $y^M = (y^\x, y^\eb)$. In deriving the   supersymmetry relation 
$\ccB_\x - \ii \ccD_\x \o = \g_\x + \ldots$, we have only used the existence of this complex structure, and not been specific about the role of individual parameters. The dimensional reduction is useful in making this identification.

The dimensional reduction proceeds perturbatively in $\ap$. We study the effective field theory for a CY background in which $H$ is $\cO(\ap)$ and $g_{mn}$ is the background metric. There is a background connection $A$ for the vector bundle $V$ whose structure group is taken to be $\eu{G}\subset E_8$.   This leaves an unbroken spacetime group given by the commutant $G = [\eu{G}, \Le_8]$ whose algebra we take to be $\Lg$. A classic example is the standard embedding in which $\eu{G} = {\rm SU}(3)$ and $\Lg = \Le_6$ but we do not restrict ourselves to this example.

The background field expansion for the ten-dimensional metric, B-field and dilaton is:
\beq\begin{split}
\dd s^2 ~&=~\big(g_{ef} + \d g_{ef}(X)\big)\, \dd X^e \otimes\dd X^f +
\big(g_{mn}(x) + \d g_{mn}(x,X)\big)\, \dd x^m\otimes \dd x^n~,\\[3pt]
B ~&=~ \d B_{ef}(X)\, \dd X^e  \dd X^f + (B_{mn}(x) + \d B_{mn}(x,X))\, \dd x^m  \dd x^n~,\\[3pt]
\Phi ~&=~  \phi_0+\vph(X) + \phi(x,X)~,
\end{split}\label{eq:ReductionAnsatz1}\eeq
where $\d g_{ef}$ is the $d{\,=\,}4$ metric fluctuation; $\d g_{mn}$ the $d{\,=\,}6$ metric fluctuation; 
$\d B_{ef}(X)$ is dual to a pseudo-scalar in spacetime, the universal axion; and the dilaton has been split into a $d{\,=\,}4$ fluctuation $\varphi(X)$ and an internal fluctuation $\phi(x,X)$ with zero-mode $\phi_0$. All small variations of background fields are regarded as quantum fluctuations.   

The decomposition of the 10D gauge field $A_M$ is partially fixed by the representation theory:
\beq
\text{Adj}(\Le_8) ~=~
(\text{Adj}(\Lg) ,{\rep 1}) \oplus_i (\rep{R}_i, \rep{r}_i)\oplus_i (\brep{R}_i,\brep{r}_i) \oplus ({\rep 1},\text{Adj}(\eu{G}))~,
\label{StandardAdjointDecomposition}\eeq
where   $\rep{R}_i$ is the matter field representation and its conjugate $\brep{R}_i$. It is obviously possible the matter fields appear in real  or psuedo-real representations $\rep{R}_i=\brep{R}_i$. We take these terms to be captured by an element of the summand $(\rep{R}_i, \rep{r}_i)\oplus(\brep{R}_i,\brep{r}_i)$  with a slight abuse of notation. In any event, the matter fields are not relevant for our calculation here.

The small fluctuations of the gauge field are given by
\beq
A   ~=~A_m(x)\, \dd x^m + \d A_e(x,X)\, \dd X^e +  \d C_m(x,X)\, \dd x^m + \d D_m(x,X)\, \dd x^m +
\d A_m(x,X)\, \dd x^m ,
\label{gaugeadjoint}\eeq
where $\d A_e(x,X)$ is the 4d gauge field in the adjoint of $E_6$; $\d C_m(x,X)$ and $\d D_m(x,X)$ are the matter fields in the $\oplus_i \rep{R}_i$ and $\oplus_i \brep{R}_i$ representation of $\Lg$; and $\d A_m(x,X)$ are $\Lg$ singlets, and is the only term relevant to the moduli space metric. We therefore drop the other fluctuations 

\subsection{Complexified metric terms $\dd s^2_g + \dd s^2_H$}
We now compute the metric explicitly by dimensionally reducing the supergravity action and identifying the coefficient the kinetic terms of the moduli fields. We start with the Lagrangians arising from the Ricci-scalar $\cL_g$ and H-field strength $\cL_H$ before including the Yang-Mills term $\cL_F$ in subsequent subsections. 
  
The expansion of the Ricci scalar to quadratic order is
\beq
\begin{split}
  \pre{10}\cR ~=  \pre{4}\cR &- \pre{4}\nabla^2 \log \det (g_6+\d g_6) -
\qrt \Big(\pre{4}\nabla \log \det (g_6 + \d g_6) \Big)^2 \cr
&- \qrt \,g^{mn}g^{pq}\del_e( \d g_{mp})\,\del^e (\d g_{nq}) + \cdots~.
\label{ricciscalar}\end{split}
\eeq
Here $\pre{4}\cR$ is the $d=4$ Ricci scalar for the metric 
$\dd s_{(4)}^2 = (g_{ef}+\d g_{ef}(X))\, \dd X^e\otimes \dd X^f$,  
while the last term will give rise to the moduli space metric. The first three terms in \eqref{ricciscalar} recombine into the $d=4$ Ricci scalar after changing to Einstein frame. To see this, we need to include the $d=4$ dilaton field defined by
$$
e^{-2\phi_4(X)} ~=~ \frac{ e^{-2\varphi(X)}}{g_s^{2}V_0} \int \! d^6 x \sqrt{g_6+\d g_6} ~.
$$ 
where $g_s = e^{\phi_0}$ is the zero-mode of the dilaton.
Then,  a Weyl transformation on the $d=4$ metric $g_{E\, ef} = e^{-2\phi_4} g_{ef}$ collapses the first three terms in  \eqref{ricciscalar} into the Einstein frame Ricci scalar. The ten-dimensional action \eqref{eq:10daction} under dimensional reduction gives 
\beq
\begin{split}
S ~=~  \frac{g_s^2}{2\k_4^2}\int d^4X \sqrt{g_{E}} \left( \pre{4}\cR_E  +\cL \right)~,
 \end{split}\label{MetricReduct}
\eeq
where  $\k_{10}^2 = V g_s^{-2}\k_4^2$, $V$ is the volume of the CY manifold and 
$\cL = \cL_g + \cL_h + \cL_S + \cL_f$ is the four-dimensional Lagrangian for the kinetic terms of the moduli fields coming from the reduction of first four terms of \eqref{eq:10daction}.  We compute each of these terms below. The remaining terms in \eqref{eq:10daction} are at least $\cO(\ap^2)$ and are ignored.

The first term  $\cL_g$ comes from the $\cR$ in \eqref{ricciscalar} and is given by
 \beq
\begin{split}
 \cL_g ~&= - \frac{1}{4V} \int d^6x \sqrt{g}\, g^{mn} g^{pq}\, \del_e (\d g_{mp})\, \del^e (\d g_{nq})~.
 \end{split}
\notag\eeq

The next term $\cL_h$ comes from  the kinetic term for $H$:
\beq
\label{SH}
\cL_H ~= -  \frac{1}{2V }\int \dd^6 x \sqrt{g}\,  |H+\d H|^2~.
\notag\eeq
In this expression, $H$ has all three legs along $\ccX$, while $\d H$ always has a leg in four-dimensional spacetime, so $H\star \d H = 0$ leaving 
\beq
\begin{split}
 \cL_H ~= -  \frac{1}{4V }\int \dd^6 x \sqrt{g} \, \d H_{emn}  \d H^{emn} ~.
\end{split}
\notag\eeq

\subsubsection{Special geometry}\label{eq:DimRedSG}
At this point it is useful to pause, and recall what happens in special geometry when the gauge connection is identified with the spin connection, $\d A {\,=\,} \d \Th$.  We do not rely on being connected to this example, but it serves the purpose of illustration for the more general case below. The only independent variations are contained within $\d g_{mn}$ and $\d B_{mn}$. Denote the $\ap$-expansion of fields as $B = B^0 + \ap B^1 + \ldots~$, $\o = \o^0 + \ap \o^1 + \ldots~$.  

A variation of the metric is 
$$
 \del_\xi (\dd s^2) ~=~2 \D_{\x\mb}^{~~~\r} g_{\r\nb}\, \dd x^\mb\otimes \dd x^\nb + 2(\del_\xi g)_{\m\nb}\,\dd x^\m \otimes \dd x^\nb~.
$$
Since $\d g_{\mb\nb}$ and $\d g_{\m\nb}$ separately solve the Lichnerowicz equation, they can be varied independently of each other, so we can assign independent parameters to these variations. The mixed component, $\d g_{\m\nb}$, is a zero-mode of the Lichnerowicz operator if and only if it is a harmonic $(1,1)$-form. Similarly, the B-field satisfies $\dd \d B^0 = 0$ and is gauge-fixed $\dd^\dag \d B^0 = 0$, so $\d B^0$ can be expanded in harmonic $(1,1)$-forms. In sum, we associate parameters to field variations as follows:
\begin{gather}
\d g_{\mb\nb} ~=~ \d z^\a  \D_{\a(\mb\nb)}~, \quad \d z^\a \in \IC, ~ {\rm for}~ \a = 1, \ldots h^{2,1}~,
\notag\\[5pt]
\d \o ~=~ \d v^r e_r, \quad \d B = \d u^r e_r, \quad \d u^r,\d v^r \in \IR,\quad e_r\in H^{1,1}(X,\IR)~, 
 {\rm for} ~ r=1,\ldots, h^{1,1}. \label{eq:SGKK}
\end{gather}
The conventional choice of gauge fixing, $\nabla^m \d g_{mn}{\,=\,}0$, implies 
$\nabla^\mb \D_{\a(\mb\nb)} {\,=\,} 0$. When this is so, each tensor $\D_{\a(\mb\nb)}$ is in one-to-one correspondence with a harmonic representative $\D_\a^{~~\r} \in H^1(X,T)$. To see this vary the \K condition $\delb \o = 0$ with respect to complex structure to give 
$$
\delb\, \ccD_\a \o^{0,2} = 0~.
$$  
As $h^{0,2} = 0$, 
$$
\ccD_\a \o^{0,2} = \ii \D_{\a[\mb\nb]} \dd x^\mb \dd x^\nb = \delb k_\a
$$ 
for some $(0,1)$-form $k_\a$. Co-closure of $\D_{\a}^{~\r}$ gives $\del^\mb \D_{\a\,\mb\nb} = \del^\mb \D_{\a\,[\mb\nb]} = 0$ and, as $\ccX$ is compact, this forces $\D_{\a[\mb\nb]} = 0$. Hence, $\D_\a^{~\r}$ is in one-to-one correspondence both with the metric variations $\d g_{\mb\nb}$ via
\beq
g^{\r\nb} \d g_{\mb\nb}~=~\d z^\a \D_{\a\mb}{}^\r~,
\label{eq:ginvdg}\eeq
and with harmonic $(2,1)$-forms $\ch_\a$ via
\beq
\chi_\a ~=~ \frac{1}{2}\O_{\r\s}^{~~~\nb}  \D_{\a\,\mb\nb}  \dd x^\r \dd x^\s \dd x^\mb~.
\label{eq:chiDefn}\eeq
The inverse of this last relation is
\beq
\D_\a^{~\m} ~=~ \frac{1}{2\norm{\O}^2} \Ob^{\m\t\r} \chi_{\a\,\t\r\,\sb} \, \dd x^\sb~.
\label{eq:DeltaDef}\eeq
We have seen these relations before in \eqref{eq:gDeltaEq} and \eqref{eq:DeltaChiEq}, though now we have specialised to the case $\ap{\,=\,}0$, for which $\D_{\a\,[\mb\nb]}{\,=\,}0$, and this has allowed us to write
\eqref{eq:ginvdg} in the given form.
It is easy to see $\chi_\a$ and $\D_\a^{~\m}$ are also $\delb$-closed and co-closed. This establishes an isomorphism $H^1(X,T) \cong H^{2,1}(X,\IC)$. 

Promoting the parameters to dynamical fields, denoted by corresponding capital letters, for example 
$u^r \to U^r(y)$,  
$\cL_g$ is
\beq
\begin{split}
\cL_g ~&= - \frac{1}{2V} \int \dd^6 x\sqrt{g} g^{\m\bar\n} g^{\r\bar\t}
\Big(  \del_e (\d g_{\nb\tb} )\del^e (\d g_{\m\r} )+ \del_e (\d g_{\m\bar \t}) \del^e (\d g_{\bar\n\r})\Big)
\\[5pt]
&= -\frac{1}{2V} \int \dd^6 x\sqrt{g}  
\left( \del_e Z^\a\del^e Z^\bb \D_{\a(\mb\nb)}  \,\D_{\bb}^{~(\mb\nb)} +\del_e \o_{\m\nb } \del^e \o^{\m\nb}\right)  \\[7pt]
&= - 2G^0_{\a\bb}\, \del_e Z^\a \del^e Z^\bb + G^0_{rs}\, \del_e V^r \del^e V^s~. 
\end{split}\label{eq:MetricReduct1}
\eeq
 where we identify the special geometry metrics
 \beq
 G_{\a\bb}^0 ~= - \frac{\ii}{V\norm{\O}^2} \int \chi_\a \star \chi_\bb~, 
 \qquad G^0_{rs} =\frac{1}{2V}  \int  e_r \star  e_s~.
\notag \eeq
 We have used the Kaluza--Klein ansatz \eqref{eq:SGKK} in writing $\del_e \o = \del_e V^r e_r$ and $\del_e (\d g_{\mb\nb}) = \del_e Z^\a \D_{\a(\mb\nb)}$  together with \eqref{eq:DeltaDef}. The $H$-field gives
\beq
 \cL_H ~=~ G^0_{rs}\, \del U^r \del U^s~.
\notag\eeq
 The complex structure moduli space automatically gives a \K moduli space metric 
 $G^0_{\a\bb}$. The \K moduli space  $\cM_K$ is also complex but the choice of complex coordinates in terms of $u^r, v^r$ is ambiguous.  The canonical choice is to associate a point $p\in \cM_K$ with a complexified form $B+\ii\o$. As $\dim_\IC \cM_K = h^{1,1}$ there are local coordinates $t^\r, t^\sb$ for $\r,\sb=1,\ldots,h^{1,1}$ to be identified. The tangent space $T_p\cM_K^\IC$ is a complex vector space, and the complex structure facilitates a splitting: $T_p\cM_K^\IC = T_p\cM_K^{1,0} \oplus T_p\cM_K^{0,1}$. 
The $e_r$ are a basis for the complexification $H^{1,1}(X,\IC)$ and so the conventional choice is 
\beq
\d B + \ii \d \o ~=~ (\d u^\r + \ii \d v^\r) e_\r \in T_p^{1,0}\cM_K \cong H^{1,1}(X,\IC) ~.
\notag\eeq
Similarly,  deformations of $B-\ii \o$ are identified as
\beq
\d B - \ii \d \o ~=~ (\d u^\sb - \ii \d v^\sb) e_\sb \in T_p^{0,1}\cM_K \cong H^{1,1}(X,\IC) ~.
\notag\eeq

The special geometry metric is then given by identifying the metric of the kinetic terms in the Lagrangian \eqref{eq:MetricReduct1}:
\beq
\dd s^2 ~=~ 2 G^0_{\a\bb}\, \dd z^\a \dd z^\bb + 2G^0_{\r\sb}\, \dd t_0^\r \dd t^\sb_0~,
\notag\eeq
where for harmonic forms $e_r$, $\chi_\a$ we can write the metrics in a form that depends only on the cohomology classes:
\beq
G^0_{\a\bb} ~= -\frac{ \int \chi_\a\, \chi_\bb}{\int \O\, \Ob}~, \qquad  
G_{\r\sb} ~=~ \half\left( \frac{1}{2V}\int e_\r \,\o^2 \right)  \left( \frac{1}{2V}\int e_\sb \,\o^2 \right)  -  \frac{1}{4V}\int \o\,e_\r \,e_\sb~.
\notag\eeq

\subsubsection{The heterotic $\ap$-corrected $\dd s^2_g+\dd s^2_H$} 
Now we proceed to the general case, including the $\ap$-correction and assuming a general choice of holomorphic semi-stable vector bundle. The Kaluza--Klein ansatz includes a correction that allows for a dependence on all parameters: 
$$
\d \o ~=~ \d v^r e_r + \ap \d y^M \ccD_M \o, \qquad  
\d H ~=~ \dd\left(\d u^r e_r + \ap \d y^M  \ccB_M \right)~.
$$
When substituted into the Ricci-scalar and $H$-field kinetic term, we identify the metric through the kinetic terms arising from $\dd s^2_g$ and $\dd s^2_H$ respectively:
\beq
\begin{split}
\dd s^2_g  &= G^0_{rs} \dd v^r \dd v^s + \ap\! \left( \frac{1}{V}\! \int\!\! \ccD_\x\o^1 \star e_s\! \right)    
\dd y^\x \dd v^s + \ap\! \left( \frac{1}{V}\! \int\! e_r \star \ccD_\eb\o^1\!\right)  \dd v^r \dd y^\eb
+ 2G^0_{\a\bb} \dd z^\a \dd z^\bb, \\[10pt]
\dd s^2_H &= G^0_{rs} \dd u^r \dd u^s   + 
\ap\!  \left( \frac{1}{V}\! \int\! \ccB^1_\x \star e_s \right) \dd y^\x\dd u^s +  
\ap\! \left( \frac{1}{V}\! \int\! e_r \star \ccB^1_\eb  \right) \dd u^r \dd y^\eb\\[7pt]
&= G^0_{rs} \dd u^r \dd u^s   + 
\ii \ap\!  \left( \frac{1}{V}\! \int\! \ccD_\x\o^1 \star e_s \right) \dd y^\x\dd u^s -  
\ii \ap\! \left( \frac{1}{V}\! \int\! e_r \star \ccD_\eb\o^1  \right) \dd u^r \dd y^\eb  \\[5pt]
&\hskip2.5cm + \ap\!\left( \frac{1}{V}\! \int\! e_r \star \g^1_\x \right) \dd u^r \dd y^\x +
\ap\!\left( \frac{1}{V}\! \int\! e_r \star \g^1_\eb \right) \dd u^r \dd y^\eb~,
\end{split}\notag
\eeq
where we have  substituted spacetime fields kinetic energy terms for metric coordinates on $\cM$ e.g. $\del_e U^r \to \dd u^r$. In the last equality, we have used the supersymmetry relation $\ccB^{1,1}_\x = \ii \ccD_\x \o^{1,1} + \g^{1,1}_\x + [\dd (\ldots)]^{1,1}$. 
We have written the special geometry metric $G^0_{rs}$, and we identify the $\ap$-correction to it:
\beq
\begin{split}
 G^{0}_{rs} ~=~ \frac{1}{2V} \int e_r \star e_s~, \qquad G^1_{\x s} ~=~ \frac{1} {2V} \int \ccD_\x \o^1 \star e_s~.
\end{split}\label{eq:Kahmetric}
\eeq
The freedom to shift by $\dd$-exact terms means we can expand $\g^1_\x$ in harmonic $(1,1)$ forms, 
$\g^1_\x = \g_\x^{1~s} e_s$ giving
$$
\left( \frac{\ap}{2V} \int e_r \star \g^1_\x \right) \dd u^r \dd y^\x ~=~ 
\left( \frac{\ap}{2V} \int e_r \star e_s \right) \g_\x^{1\,s}\, \dd u^r  \dd y^\x ~=~
\ap G_{rs}^0 \g_\x^{1\,s}\, \dd u^r  \dd y^\x~.
$$
Adding $\dd s^2_g$ and $\dd s^2_H$ together
\beq
\begin{split}
 \dd s^2_g + \dd s^2_H ~&=~ 2G^0_{\a\bb}\, \dd z^\a \dd z^\bb + 
 G^{0}_{rs}\Big(\dd v^r \dd v^s +  \dd u^r \dd u^s + \ap \g^{1\,r}_\x \dd y^\x \dd u^s + \ap \g^{1\,r}_\eb \dd y^\eb \dd u^s  \Big) \\
 &\hskip3cm + 2\ii \ap G^1_{\x s}\, \dd y^\x (\dd u^s - \ii \dd v^s ) - 
 2\ii \ap G^1_{ r\eb}\, (  \dd u^r +\ii \dd v^r)\,\dd y^\eb~   \\[8pt]
 &=~ 2G^0_{\a\bb}\, \dd z^\a \dd z^\bb +
 G^{0}_{rs}\Big(\dd u^r + \ii \dd v^r + \ap \g_\eb^{1\,r}\,\dd y^\eb \Big)
 \Big( \dd u^s -\ii \dd v^s  + \ap \g_\x^{1\,s}\,\dd y^\x\Big)\cr
 &\hskip3cm + 2\ap G^1_{\x s} \dd y^\x (\dd v^s + \ii \dd u^s) + 
 2\ap G^1_{ r\eb} (\dd v^r - \ii \dd u^r)\,\dd y^\eb~.\\
 \end{split}\notag
\eeq
The penultimate line indicates the complex coordinates on the parameter space $\cM$ are modified at first order in $\ap$. We can view this as a change in special geometry complex~structure: 
\beq
\dd t^\r ~=~  \dd u^\r +\ii \dd v^\r + \ap \g^{1\,\r}_\eb\, \dd y^\eb  ~=~ 
\dd t_0^\r + \ap \g^{1\,\r}_\eb\, \dd y^\eb.
\notag
\eeq
Indeed, $\g_\eb^{~\r} \,\dd y^\eb$ is a  function of only parameters and is exactly a $(0,1)$-form on $\cM$ valued in $T\cM$. 
In heterotic geometry, there is a natural modification in complex structure at $\ap$, and in the new coordinates the metric is
\beq
\begin{split}
 \dd s^2_g + \dd s^2_H ~&=~  2G^0_{\a\bb}\, \dd z^\a \dd z^\bb + 2G^{0}_{\r\sb}\, \dd t^\r \dd t^\sb +  
 2\ap G^1_{\x \sb}\, \dd y^\x \dd t^\sb + 2\ap G^1_{ \r\eb}\, \dd t^\r \dd y^\eb~,   \\[4pt]
\end{split}
\notag\eeq
where $G^0_{\r\sb} \cong 2 G^0_{rs}$ and $\dd y^\x = (\dd t^\r, \dd z^\a, \dd w^i)$ so that $\ccD_\x \o^0 {\,=\,} (-\ii /2)\, e_\x$ with the understanding that $e_\x = 0$ unless $\x=\r=1,\ldots,h^{1,1}$ is a \K parameter index. In fact, we can simplify the metric further
\beq
\begin{split}
 \dd s^2_g + \dd s^2_H ~&=~ 2G^0_{\a\bb}\, \dd z^\a \dd z^\bb + 2G^K_{\x\eb}\, \dd y^\x \dd y^\eb~,~~~
 {\rm where}~~~ G^K_{\x\eb} ~=~ \frac{1}{V} \int \ccD_\x \o \star \ccD_\eb \o~.
\end{split}\label{eq:metricDimRed}
\eeq

It is useful to flesh out the change in coordinates. The basis of 1-forms for the co-tangent space $T^*\cM$ and their inverse are related as
\beq
\begin{split}
 \dd t^\r ~&=~  \dd t_0^\r + \ap \g^{1\,\r}_\eb\, \dd y^\eb~,\qquad 
 \diff{t^\r} ~=~  \diff{t_0^\r} - \ap \g^{1\,\eb}_\r\, \diff{y^\eb}~.
\end{split}\notag
\eeq
The remaining coordinates for complex structure and the bundle are unchanged as the (1,1)-form $\g_\eb$ is harmonic and so $\g^{~i}_\eb = \g^{~\a}_\eb = 0$. 
The transformation law is  
$$
\ccB_\x^{\rm new} =~ \ccB_\x^{\rm old} - \ap\g^1_\x{}^\sb\, \ccB_\sb^{\rm old}~.
$$
The change of coordinates is viewed perturbatively in $\ap$; we find that the $(1,1)$-component~is
\beq
\begin{split}
 \ccB_\x^{\rm new} - \ii \ccD_\x \o^{\rm new} &=~ \ccB_\x^{\rm old} - \ii \ccD_\x \o^{\rm old} - \ap\g^{1\,\sb}_\x\, (\ccB_\sb^{\rm old} - \ii \ccD_\sb \o^{\rm old} ) \\[3pt]
 &=~ \ap \g^{1\,\sb}_\x e_\sb -   \ap\g^{1\,\sb}_\x\,  e_\sb\\[3pt]
 &=~ 0~.
\end{split}
\notag\eeq
We view this equation as implicitly projected onto its $(1,1)$-component and $\ccB^{old}_\sb - \ii \ccD_\sb \o^{old} =  e_\sb + \cO(\ap)$. Recall $\ccB_\x$ in \eqref{eq:curlyBdef} is defined up to a $\dd$-exact piece, and so we can absorb it into the definition of $\ccB_\x$, if we wish. Either way it does not contribute to the metric.  We conclude that
\beq
\label{eq:NewSusyRelation}
\left( \ccB_\x^{\rm new} - \ii \ccD_\x \o^{\rm new}\right)^{1,1} ~=~ 0~.
\notag\eeq
As $\g_\x$ is a $(1,1)$-form the coordinate change does not influence the $(0,2)$-component of $\ccB_\x- \ii \ccD_\x\o$. 

\subsection{Yang-Mills term $\dd s^2_F$}
We now turn to the Yang-Mills term
\beq\notag
\cL_{F} ~= -\frac{\ap}{4V}\int_\ccX \dd^6 x \sqrt{g}\, \tr |\d F|^2 ~; \qquad |\d F|^2 = \frac{1}{2} F_{MN} F^{MN}~.
\eeq
The gauge connection $A$ and spin connection $\Th$ both depend on parameters 
\begin{align}
\d \A ~&=~ \d y^\xi \,\ccD_\xi \cA~, & \d \A^\dag ~&=~ \d y^\eb\, \ccD_\eb \A^\dag, \notag\\
\d \vth ~&=~ \d y^\xi \,\ccD_\xi \vth~, & \d \vth^\dag ~&=~\d y^\eb\, \ccD_\eb \vth^\dag~.\label{bundlefluct1}
\end{align}
We take $\d \A$ and $\d \vth$ vary independently of each other, and each are taken to depend in a general way on all parameters $y^\x, y^\eb$.

We can dimensionally reduce just the Yang-Mills term, and at the end of the calculation use the parity symmetry to insert the corresponding expression for the spin connection. To that end, substituting \eqref{bundlefluct1} into the fluctuation of the field strength $F\to F + \d F$ we find
\beq\begin{split}
\d F ~&=~\dd(\d \A-\d \A^\dag) + \big\{\cA -\cA^\dag, \d \A-\d \A^\dag\big\}\\[8pt]
&=~  \dd Y^\xi\,\ccD_\xi \cA - \dd Y^{\eb}\ccD_\eb \A^\dag   + \ldots,
\end{split}\notag\eeq
where we have kept only terms that will contribute to the moduli space metric. Substituting into the Yang-Mills term:
\beq
\begin{split}
\tr |F+\d F|^2 ~&=~  \tr |F|^2 + \half \tr \d F_{MN} \d F^{MN} \\[5pt]
&=~  \tr |F|^2+ \tr\big( \d F_{e\m} \d F^{e\m}\big) + \tr\big( \d F_{e\bar\n} \d F^{e\bar\n}\big) +\cdots
\\[8pt]
&=~  \tr |F|^2 - 2g^{\m\bar\n}\, 
\del_e Y^\xi \del^e Y^{\eb}\,\tr\big( \ccD_\xi\cA_{\bar\n}\, \ccD_{\eb}\cA^\dag_\m \big) ~,
\end{split}\notag
\eeq
where the first term is a constant and to be dropped, and we have used
$\tr\big(\d F_{e\m}\, \d F^{e\m}\big) = \tr \big(\d F_{e\bar\n}\, \d F^{e\bar\n}\big)$. On substituting and including the spin connection we find:
\beq
\begin{split}
 \cL_F ~&=- 2G^\YM_{\xi\eb}\, \del_e Y^\xi \del^e Y^\eb~;~~~ 
 G^\YM_{\xi\eb} ~=~ \frac{\ii\ap}{8V}\int\o^2 
\Big( \tr\big(\ccD_{\xi} \A\, \ccD_{\eb}\A^\dag\big) - \tr\big(\ccD_{\xi} \vth\, \ccD_{\eb}\vth^\dag\big)\Big)~.
\end{split}\notag
\eeq
\vspace{-5pt}
The metric is identified as $\dd s^2_F = G^\YM_{\xi\eb} \dd y^\x \dd y^\eb$. 
\subsection{The final result: the $\ap$-corrected heterotic moduli metric}
Putting $\dd s^2_F, \dd s^2_g, \dd s^2_H$ together, we find a compact expression for the moduli space metric:
\beq
\begin{split}
\dd s^2 ~&=~ 2G^0_{\a\bb}\, \dd z^\a \dd z^\bb + 2G_{\x\eb}\, \dd y^\x \dd y^\eb~,
\end{split}\label{eq:FinalMetric1}
\eeq
where the holomorphic \K coordinates are identified as
\beq
\begin{split}
 \dd t^\r ~&=~  \dd t_0^\r + \ap \g^{1\,\r}_\eb\,\dd y^\eb~,
\end{split}
\notag\eeq
the complex structure coordinates $\dd z^\a$ are deformations that preserve the Atiyah condition as per the discussion in section X, and $\dd w^i$ are deformations of the gauge connection preserving HYM for fixed $\ccX$.  The metric components are
\beq
\begin{split}
 G_{\x\eb} ~&=~ \frac{1}{V} \int \ccD_\x \o \star \ccD_\eb \o + 
 \frac{\ii\ap}{8V}\int\o^2 \Big( \tr\big(\ccD_{\xi} \A\, \ccD_{\eb}\A^\dag\big) - 
 \tr\big(\ccD_{\xi} \vth\, \ccD_{\eb}\vth^\dag\big)\Big)~,\\[10pt]
G_{\a\bb} ~&= -\frac{ \displaystyle \int \chi_\a\, \chi_\bb}{\displaystyle \int \O\, \Ob}~,
\end{split}\label{eq:FinalMetric2}
\eeq
where $z^\a\oplus w^i$ are the holomorphic parameters corresponding to the decomposition in \eqref{AtiyahCohomology}, and $t^\r$ holomorphic parameters for the $\ap$-corrected \K parameters. 
This metric precisely matches the metric arising from the \K potential in the previous section, and so is manifestly \K as required by supersymmetry. 

There is an additional term in the metric coming from the universal axio-dilaton. It is special as it is not coupled to any of the other fields, to this order in string perturbation theory, and the result is well-known
\beqnn
\cL_S ~=~ 2\,\frac{ \del_e S\, \del^e \overline{S} }{S - \overline{S}}~.
\eeqnn

\newpage
\section{The universal geometry of heterotic moduli}\label{sec:UnivBundle}
The central results of this paper were originally derived with the goal of finding the natural \K metric on the moduli of heterotic structures. In this process, we constructed, in  \sref{sec:params}, the holotypical derivatives of various fields including $\o$, $\A$ and $B$, in order to describe deformations of these fields in a gauge covariant manner. Supersymmetry imposes relations between the first and second order derivatives, some of which were explored in  \sref{sect:susyrelations}. In isolation, the meaning of these derivatives and identities is obscure, beyond the technical need to construct a metric. Remarkably, however, we have found they have a natural geometric interpretation when viewed in the context of what is known as the universal bundle. 

In retrospect, the idea is  simple: instead of treating the parameter space and heterotic structures separately, we should study the geometry of the total space of the family of heterotic structures. The family, denoted $\ccH$, is assumed to have a local product structure, so that $\ccH$ is a fibre bundle $\ccH\to \cM$, with the fibres being the heterotic structures. A heterotic structure, recall, is the total space of the stable holomorphic bundle $\ccE\to \ccX$ together with a gauge invariant three-form $H$ defined on $\ccX$. Just as in \sref{sec:params} it was natural to extend gauge transformations to depend on the coordinates of $\cM$ and  $\ccX$, we extend the gauge field and three-form to live on the family $\ccH$. In the context of the geometry of $\ccH$, the quantities constructed in \sref{sec:params}-\sref{sect:susyrelations} take a new, simpler, geometric meaning. The holotypical derivative is interpreted as a covariant derivative on $\ccH$, defining parallel transport of heterotic structures. What is interesting, is if we extend supersymmetry to the family~$\ccH$. That is, take 
$\ccH$ to be a complex manifold with a corresponding hermitian form. Then make the ansatz that the hermitian form and the three-form are related via an identity that naturally extends $H= \ii (\del-\delb) \o$. We can then show that this relation, together with the Bianchi identities for the gauge field and the three-form, captures all of the identities needed to derive the moduli space metric. The gauge freedom implicit in the construct facilitates a natural construction of the covariant derivatives $D_\x \A$, the holotypical derivatives $\ccD_\x B$ and even of intricate objects such as $\ccB_\x$, associated to gauge invariant deformations of $H$. This idea of geometrising moduli spaces is reminiscent of F-theory and generalised geometry and it would be intriguing to explore these connections further. Indeed, in forthcoming work \cite{CdOMcO:2016a} we describe this geometry in detail, and how it summarises efficiently many of our results. 

We illustrate the idea in the restricted case of a fixed $\ccX$. In that case, we are studying the moduli space of connections $A$, with parameters $w^I$, satisfying the hermitian Yang-Mills equation over a fixed $\ccX$. With the notation
$$
D~=~\dd w^I D_I~~~,~~~\widetilde{\dd}~=~\dd w^I \partial_I~~~,~~~\L~=~\dd w^I \L_I~.
$$
In this notation the covariant derivative of $A$ can be written
$$
D A~=~\widetilde{\dd} A + \dd\L + \{A,\L\}~.
$$
The form of this equation already suggests a connection with a field strength.  Indeed, if we define a connection $\IA=A+\L$  on $\ccH$ and study the field strength $\IF$ of $\IA$ we find
$$
\IF~=~\Id\IA + \IA ^2~=~(\dd A+A^2) + (\widetilde{\dd}\L +\L^2) + D A
$$
 where $\Id=\dd+\widetilde{\dd}$. What this already teaches us is that the covariant derivative we worked hard to derive previously, corresponds to the form $\IF$ with mixed indices between the parameter space $\cM$ and  $\ccX$. In the broader context of the heterotic string we show how this is extended to the holotypical derivative and how it is interpreted as a connection. 

If we impose that $\IF$ a holomorphic field strength, then the vanishing of the $(0,2)$-component amounts to 
$\ccD_\jb \A = 0$ -- that is, we find a geometric condition for $\A$ admitting a holomorphic dependence on parameters. 

Consider now the Bianchi identity for the field strength $\IF$:
$$
\Id_\IA {\IF}~=~\Id\IF + [\IA, \IF]~=~0~.
$$ 
 The mixed components of this equation give precisely two of the identities that we derived previously. The first comes from considering the $\dd w^I \dd x^m \dd x^n$ components
$$
D_I F - \dd_A(D_I A) = 0~,
$$
which is exactly \eqref{DF}; and a second comes from considering the $\dd w^I \dd w^J \dd x^n$
components:
\beq
\big[ D_I, D_J \big] A~=- \dd_A \IF_{IJ}~.
\label{eq:DerivCommA}\eeq

The salient point is that geometrising our algebraic structures is a powerful way of viewing the moduli space. 
\section*{Acknowledgements}
\vskip-10pt
It is a pleasure to acknowledge interesting conversations with Nigel~Hitchin, Zhentao~Lu, Ilarion~Melnikov, 
Mudumbai~Narasimhan, Andy~Royston, and Eirik~Svanes. We wish to  acknowledge the hospitality and support of the ICTP, Trieste, where part of this work was done. PC and XD are  supported by EPSRC grant BKRWDM00. JM is supported by STFC grant ST/L000490/1. 
\newpage
\appendix
\section{Hodge theory for Calabi-Yau manifolds}\label{s:bundlefacts}
\vskip-5pt
Some useful results from the differential geometry and Hodge theory. 

Define the Hodge dual of a $p$-form $A_p$ on an $n$-dimensional manifold to be
$$
\star A_p ~=~ \frac{1}{p!(n-p)!} \eta^{m_1\ldots m_p}{}_{n_1\ldots n_{n-p}} A_{m_1\ldots m_p} dx^{n_1} \ldots dx^{n_{n-p}}~.
$$
where $\eta_{m_1\ldots m_n} = g^{1/2}\e_{m_1\ldots m_n}$, with $g$ is the determinant of the metric on $\ccX$ and $\e_{m_1\ldots m_n}$ is the totally antisymmetric Levi-Civita symbol. The volume form is defined as
\beq
\star 1 = \frac{1}{n!} \eta_{m_1\ldots m_n} dx^{m_1}\ldots dx^{m_n}, \quad V_n = \int \star 1~.
\label{eq:RealVol}\eeq
The inner product of two $p$-forms is then given by
$$
A_p \w \star B_p ~=~ \frac{1}{p!} A_{m_1 \ldots m_p}B^{m_1\ldots m_p} \star 1.
$$

Due to the type structure of the epsilon tensor, the Hodge dual acts
$$
\star\,: \O^{r,s}(X) ~\longrightarrow~ \O^{m-s,m-r}(X).
$$
We can define an inner product on $\a,\b\in\O^{r,s}(X)$ as
\beq
(\a,\b) = \int \a \starb \b,
\notag\eeq
where $\starb: \O^{r,s}(X) \to \O^{m-r,m-s}(X)$. It is related to the usual Hodge dual as
$$
\starb \b = {\overline{ \star \b}} = \star {\overline \b}.
$$
We will need the adjoint operator $\delb^\dag: \O^{r,s} \to \O^{r,s-1}$. In terms of $\starb$:
$$
\delb^\dag  = - \starb \delb\starb = -\overline{ (\star\,\del\,\star\,)}
$$
We will often need to compute the dual of $(1,1)$-forms $F$:
\beq
\star F ~=~ \frac{1}{2} F^{\m\nb} \o_{\m\nb} \, \o^2 -  F \,\o~.
\label{eq:Starmap1}
\eeq
We derive this using $\frac{1}{3!} \o^3 = \star 1$, $\o_{\m\nb} = \ii g_{\m\nb}$ and $\o$ is $(1,1)$ and so the result applies when $\o$ is a $(1,1)$ Hermitian form. Notice that when $F=\o$:
$$
\star\,\o ~=~ \half \o^2.
$$
The Hermitian form is $\delb$-closed if and only if it is $\delb$-coclosed. 

If $\o,F$ are $\delb$-harmonic then acting with $\delb$ on \eqref{eq:Starmap1} gives us that 
$F^{\m\nb} \o_{\m\nb}$ is a constant, so
$$
F^{\m\nb}\o_{\m\nb} = \frac{1}{V} \int F \star \o = \frac{1}{2V} \int F \o^2~.
$$
The image of the $\star$-map depends only on the cohomology classes of $\o$ and $F$:
\beq
\star\, F = \frac{1}{4V}\left(\int F \o^2\right) \o^2 - F \o.
\label{eq:Starmap2}
\eeq

On a complex $p$-manifold we have
\beq
\star 1 ~=~\frac{g^\frac12} {(p!)^2} \begin{cases} 
 \ii\,\e_{\m_1\ldots \m_p} \e_{\nb_1\ldots \nb_p} \dd z^{\m_1}\ldots \dd z^{\m_p} 
 \dd z^{\nb_1}\ldots \dd z^{\nb_p}\qquad p~{\rm odd},\\[5pt]
 \e_{\m_1\ldots \m_p} \e_{\nb_1\ldots \nb_p} \dd z^{\m_1}\ldots \dd z^{\m_p} 
 \dd z^{\nb_1}\ldots \dd z^{\nb_p}, \qquad p~{\rm even},
\end{cases}
\label{eq:ComplexVol}
\eeq
We fix the coefficient by demanding \eqref{eq:ComplexVol} reduce to \eqref{eq:RealVol} when $\dd z^\m = \dd x^{\m} + \ii \dd y^{\m}$ with  orientation defined as $\dd x^1 \dd y^1 \dd x^2 \dd y^2 \ldots \dd x^p \dd y^p$. 

For us $p=3$, 
\beq
\star 1 =\frac{\sqrt{g}} {(3!)^2} \ii\e_{\m_1\ldots \m_p} \e_{\nb_1\ldots \nb_p} \dd z^{\m_1} \dd z^{\m_2} \ldots\dd z^{\nb_1}\ldots \dd z^{\nb_p}.
\label{eq:3foldComplexVol}
\eeq
On the other hand $\star 1 = \frac{1}{3!} \o^3$ and so using $\o_{\m\nb} = \ii g_{\m\nb}$, the exterior forms automatically project onto the antisymmetric combination of indices:
\beq
\begin{split}
 \frac{1}{3!} \o^3 ~&=~ \frac{1}{3!}\, \o_{[\m_1|\nb_1} \o_{|\m_2|\nb_2} \o_{\m_3]\nb_3} \, \dd z^{\m_1} \dd z^{\nb_1} \ldots \dd z^{\nb_3}\\[5pt]
&=~\frac{\ii}{3!}\, g_{[\m_1|\nb_1} g_{|\m_2|\nb_2} g_{\m_3]\nb_3} \, \dd z^{\m_1} \dd z^{\m_2} \dd z^{\m_3} \dd z^{\nb_1}\dd z^{\nb_2} \dd z^{\nb_3}.
\end{split}\notag
\eeq
Note that $g_{[\m_1|\nb_1} g_{|\m_2|\nb_2} g_{\m_3]\nb_3}$, being antisymmetric in $\m_i$, means that it is automatically antisymmetric in $\nb$. 
We identify
$$
\frac{\sqrt{g}} {3!} \e_{\m_1\m_2 \m_3} \e_{\nb_1\nb_2 \nb_3} ~=~  g_{[\m_1|\nb_1} g_{|\m_2|\nb_2} g_{\m_3]\nb_3},
$$
which can be checked to be consistent with the determinant of a hermitian metric:
\beq
\sqrt{g}  = \frac{1}{3!} \e^{\m_1\m_2 \m_3} \e^{\nb_1\nb_2 \nb_3}  g_{[\m_1|\nb_1} g_{|\m_2|\nb_2} g_{\m_3]\nb_3},
\notag\eeq
where $\e^{123} = 1$. Now consider a $(1,1)$-form $F= F_{\m\nb} \dd z^\m \dd z^\nb$. Its hodge dual
\beq
\star F ~= -\ii \frac{ \sqrt{g}}{(2!)^2}  g^{\r\bar\n_1}\e_{\nb_1\nb_2\nb_3} g^{\m_1\sb}\e_{\m_1\m_2\m_3} F_{\r\sb}  \, \dd z^{\m_2} \dd z^{\m_3} \dd z^{\nb_2} \dd z^{\nb_3}.
\notag\eeq
We fix the sign of the dual by demanding $\star \o = \half \o^2$. 

Expanding in full detail:
\beq
\begin{split}
 \star F ~&=~  \frac{3\ii }{2} \frac{\sqrt{g}}{3!}  \e_{\nb_1\nb_2\nb_3} \e_{\m_1\m_2\m_3} F^{\m_1\nb_1}  \, \dd z^{\m_2} \dd z^{\m_3} \dd z^{\nb_2} \dd z^{\nb_3}\\[3pt]
&=~  \frac{\ii }{4} \left(2g_{\m_1\nb_1} g_{\m_2\nb_2} g_{\m_3\nb_3} + 4g_{\m_2 \nb_1} g_{\m_3\nb_2} g_{\m_1\nb_3}  \right) F^{\m_1\nb_1}\, \dd z^{\m_2} \dd z^{\m_3} \dd z^{\nb_2} \dd z^{\nb_3}\\[3pt]
&=~ \frac{1}{2} F^{\m\nb} \o_{\m\nb}\, \o^2 - F \o.
\end{split}
\notag\eeq
Thus we end up with the result \eqref{eq:Starmap1}. In this calculation, we have used $\o_{\m\nb} = \ii g_{\m\nb}$; antisymmetry in $\m$ implies antisymmetry in $\nb$; and 
\beq
\sqrt{g} g^{\m_1\nb_1} \e_{\m_1\m_2\m_3}\, \e_{\nb_1\nb_2\nb_3} ~=~ 
g_{\m_2\nb_2}g_{\m_3\nb_3} - g_{\m_3\nb_2}g_{\m_2\nb_3}~.
\notag\eeq
\vskip3cm
\baselineskip=8.5pt

\providecommand{\href}[2]{#2}\begingroup\raggedright\endgroup
\end{document}